\documentclass[rmp,reprint,aps,amsmath,amssymb,superscriptaddress]{revtex4-2}

\usepackage[colorlinks=true,linkcolor=red,citecolor=blue,urlcolor=blue]{hyperref}
\usepackage{orcidlink}

\usepackage{dcolumn,graphicx,siunitx}

\newcommand{\be}{\begin{equation}}
\newcommand{\ee}{\end{equation}}

\newcommand{\bea}{\begin{eqnarray}}
\newcommand{\eea}{\end{eqnarray}}

\newcommand{\kb}{k} 
\newcommand{\rn}{\rho}  

\newcommand{\mt}{m_{\mathrm{s}}} 

\newcommand{\ud}{U_{\mathrm{d}}} 

\newcommand{\gtot}{\Gamma_{\mathrm{tot}}}
\newcommand{\gloss}{\Gamma_{\mathrm{loss}}}

\newcommand{\stot}{\sigma_{\mathrm{tot}}}
\newcommand{\sloss}{\sigma_{\mathrm{loss}}}

\newcommand{\thetamin}{\theta_{U}}
\newcommand{\svloss}{\langle v_{\rm rel}\sigma_{\rm{loss}}(E,U)\rangle}
\newcommand{\svtot}{\langle v_{\rm rel}\sigma_{\rm{tot}}(E)\rangle}

\newcommand{\vp}{v_{\rm p}}
\newcommand{\svtotCsix}{\langle v_{\rm rel}\sigma_{\rm tot}(E) \rangle_{\rm{C6}}}

\begin{document}

\title{Roadmap on Quantum Sensors for Vacuum Metrology}

\author{D. S. Barker\orcidlink{0000-0002-4614-5833}}
\affiliation{National Institute of Standards and Technology, 100 Bureau Drive, 
Gaithersburg, Maryland 20899, USA}

\author{J. L. Booth\orcidlink{0000-0001-9102-3141}}
\affiliation{Physics Department, British Columbia Institute of Technology, 
 3700 Willingdon Avenue, Burnaby,  BC V5G 3H2, Canada}

\author{S. P. Eckel\orcidlink{0000-0002-8887-0320}}
\affiliation{National Institute of Standards and Technology, 100 Bureau Drive, 
Gaithersburg, Maryland 20899, USA}

\author{J. Grosse\orcidlink{0000-0002-3942-8819}}
\affiliation{Center of Applied Space Technology and Microgravity (ZARM), University of Bremen, Am Fallturm 2, 28359 Bremen, Germany}

\author{K. Jousten\orcidlink{0000-0003-1173-1514}}
\affiliation{Department of Heat and Vacuum Metrology, Physikalisch-Technische Bundesanstalt, Abbestr.2-12, 10587 Berlin, Germany}

\author{J. K\l{}os\orcidlink{0000-0002-7407-303X}}
\affiliation{Joint Quantum Institute, Department of Physics, University of Maryland, College Park, Maryland 20742, USA}

\author{R. V. Krems}
\affiliation{Department of Chemistry and Stewart Blusson Quantum Matter Institute, University of British Columbia, Vancouver, BC V6T 1Z1, Canada} 

\author{K. W. Madison\orcidlink{0000-0003-0331-3267}}
\email[Corresponding author: ]{madison@phas.ubc.ca}
\affiliation{Department of Physics and Astronomy, University of British Columbia, Vancouver, BC V6T 1Z1, Canada}

\author{T. Rubin}
\affiliation{Department of Heat and Vacuum Metrology, Physikalisch-Technische Bundesanstalt, Abbestr.2-12, 10587 Berlin, Germany}

\author{E. Tiesinga\orcidlink{0000-0003-0192-5585}}
\email[Corresponding author: ]{eite.tiesinga@nist.gov}
\affiliation{National Institute of Standards and Technology,
100 Bureau Drive, Gaithersburg, Maryland 20899, USA}
\affiliation{Joint Quantum Institute, Department of Physics, University of Maryland, College Park, Maryland 20742, USA}

\begin{abstract}
We describe how laser-cooled and trapped alkali-metal atoms have been used as pressure sensors in the high- and ultra-high-vacuum regimes operating with relative uncertainties of just over a percent at pressures from a few nPa and up to a few tens of $\mu$Pa. 
These sensors can serve as primary standards of pressure as the determination of pressure requires only knowledge of collision cross sections between atoms and molecules, eliminating the need for calibration with other devices. The collision cross sections can be measured independently or computed with rigorous quantum methods at percent relative accuracies. 
We argue that this technology has advanced sufficiently to provide an alternative to the existing, expensive pressure standard in the ultra-high vacuum regime, the orifice flow standard. Experiments with trapped atoms can be used to calibrate cheaper pressure sensors (such as devices based on ionization gauges). However, the technology has not yet matured enough to be of use in other research fields because of the complexity and cost of the experiments. We propose a roadmap for this new pressure standard for vacuum metrology.  
\end{abstract}

\maketitle

\tableofcontents

\section{Introduction}
\label{sec:introduction}

For at least the past 100 years, the Pascal, the unit of pressure in the International System of Units (SI), has been realized by mechanical means through a force per unit area.
Specifically, the force per unit area on the surface of a mercury column has been the primary standard for pressures $p$ for low or rough vacuum between 100~Pa and atmospheric pressure at approximately 100~kPa~\cite{Jousten_2017}.
The force on the surface of a piston embedded in a cylinder with a small gap  between the piston and the cylinder walls, a so-called piston gauge or pressure balance, realizes pressure from near-atmospheric pressure up to 7~MPa~\cite{Jousten_2017}.
Today, both methods achieve relative standard uncertainties $u(p)/p$ of the order of $10^{-6}$.

\begin{figure}
   \includegraphics[width=\linewidth]{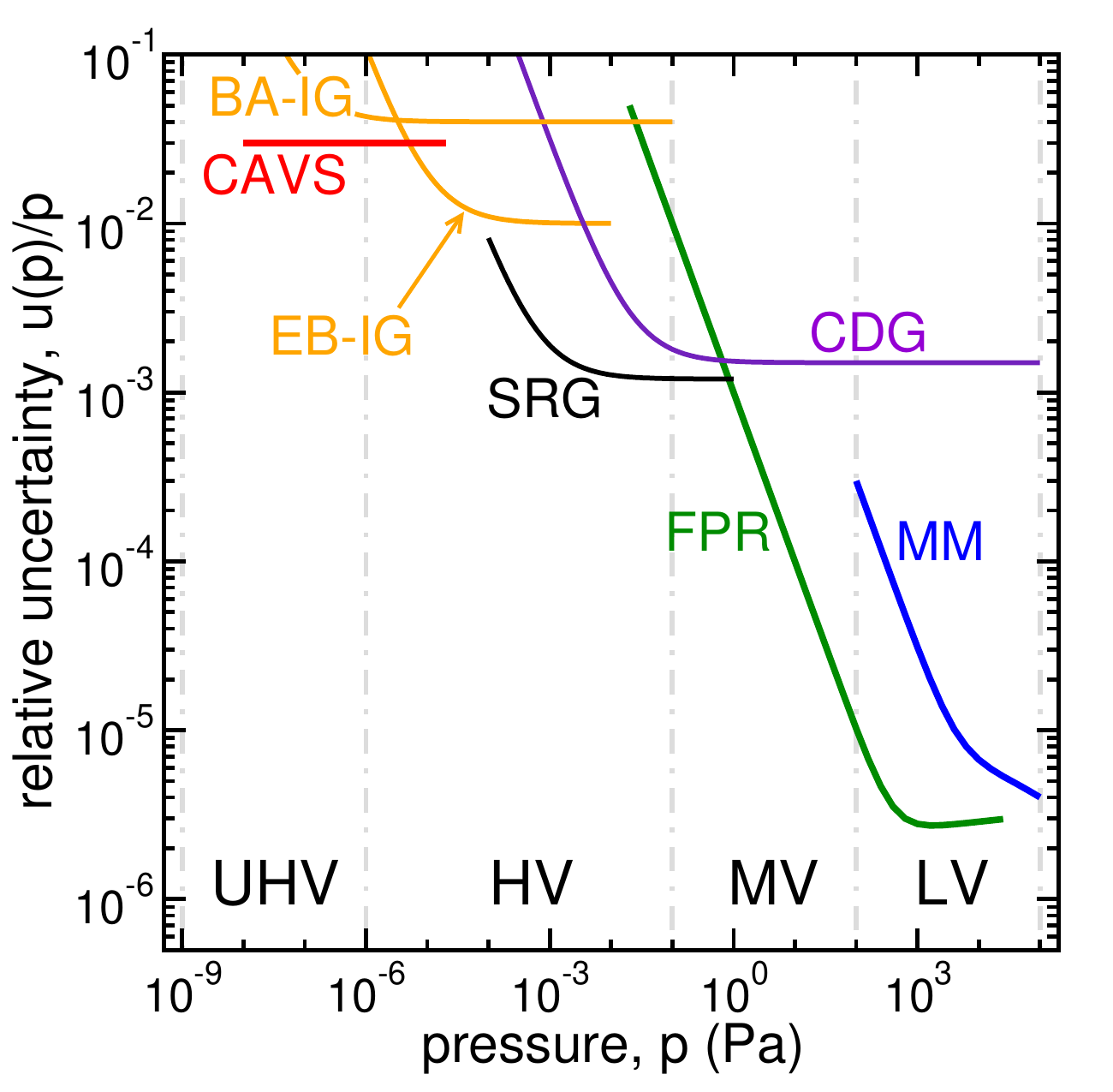} 
   \caption{
        Relative standard uncertainties $u(p)/p$ as functions of pressure $p$ for  pressure-sensing devices or methods based on force per area as well as on measuring number densities. 
        Near the bottom of the figure, we define the pressure ranges separated by dot-dashed gray lines for  ultra-high-vacuum (UHV), high-vacuum (HV), medium-vacuum (MV), and low-vacuum (LV). Extreme-high vacuums (XHV) have $p<10^{-9}$~Pa (not shown). Atmospheric pressure corresponds to $10^5$ Pa.
        Colored solid curves show the current relative uncertainties of mercury manometers (MM, blue), capacitance diaphragm gauges (CDG, purple), Fabry-P\'erot refractometry (FPR, green), spinning rotor gauges (SRG, black), Bayard-Alpert ionization gauges (BA-IG, orange), electron-beam ionization gauges (EB-IG, orange), and, finally, cold-atom vacuum sensors (CAVS, red), the subject of this article. 
        The first two devices in this list rely on force measurements, 
        the remaining devices measure number densities.
        The devices operate in different but overlapping vacuum regimes.
       } 
        \label{fig:CAVS_and_other_density-based_methods}
\end{figure}

Both mercury manometers and piston gauges serve, together with other techniques that generate known pressures such as static expansion~\cite{poulter_calibration_1977, jitschin_pressures_1990} and dynamic expansion~\cite{Tilford1988, Jousten1999, Jousten2002, Eckel2022, Barker2022}, as standards for the calibration of a host of pressure sensors.
In Fig.~\ref{fig:CAVS_and_other_density-based_methods}, we show the total relative uncertainty $u(p)/p$ with calibration for several, metrological-quality vacuum pressure sensors.
Among them are capacitance diaphragm gauges (CDGs), spinning rotor gauges (SRGs)~\cite{Fremerey1985, Comsa1980}, Bayard-Alpert ionization gauges (BA-IGs)~\cite{Bayard1950}, and, most recently, the electron-beam ionization gauge (EB-IG)~\cite{Jenninger2021,Jousten2021,Jousten2023}.

With the revision of the International System of Units (SI) in 2019, new realizations of the base and derived units have emerged, including the pascal for pressure.
Common to these new realizations is that the pascal is realized by measuring gas number density $\rn$ and temperature $T$.
The ideal gas law then gives pressure 
\be
p=\rn \kb T\,,
\label{eq:idealgaslaw}
\ee
where $\kb$ is the Boltzmann constant, a constant with no uncertainty in the revised SI.
Importantly, these emerging realizations operate in modes that {\it sense} pressure rather than merely generating it.
Thus, these realizations have the potential  to be both sensors and standards, potentially eliminating the need  for calibration and the standards required for calibration.

Over the past ten years and anticipating the redefinition of the SI \cite{Jousten_2017}, the measurement of $\rn$ using Fabri-P\'erot refractometry has become  more accurate than mechanical force-based realizations of the pascal for ${p>1}$~Pa as shown in Fig.~\ref{fig:CAVS_and_other_density-based_methods}.
In this technique, gas number density $\rn$ is determined by measuring the refractive index of the gas, which changes linearly with number density via the Lorentz-Lorenz equation and its generalizations~\cite{egan2015, egan2016, Jousten_2017, Yang2021}. 
If the proportionality constant, {\it i.e.} the dynamic polarizability of the gas, is known from {\it ab-initio} quantum physical calculations and the refractive index is measured at the vacuum system's base pressure (the system's lowest achievable pressure), number densities measured with refractometry are primary.
Refractrometry is most accurate between 1~kPa and 100~kPa with relative uncertainties just above $10^{-6}$.

With the advent of laser cooling and trapping of atoms in the 1980s, a second primary sensor based
on measuring gas number densities that excels in the high and ultra-high vacuum regimes
has become possible. Trapped ultracold atoms can serve as a sensor of background-gas number density at the trap location in the vacuum. 

\begin{figure}
   \includegraphics[width=\linewidth]{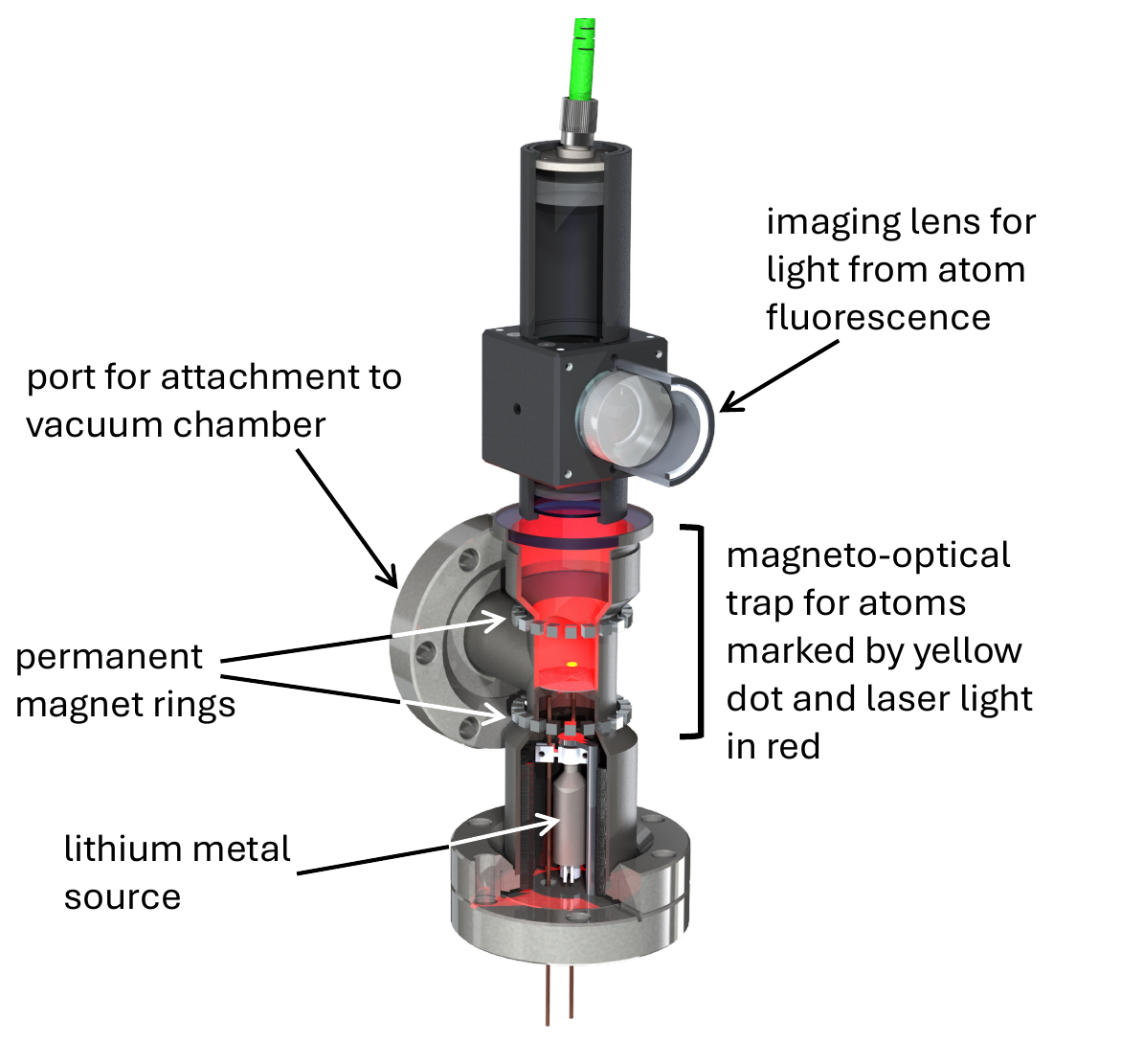} 
        \caption{Image of a lithium-based portable cold-atom vacuum sensor (p-CAVS) 
           {
           with its port to a vacuum chamber. The device was built at the US National Institute of Standards and Technology (NIST)~\cite{Ehinger2022}.
        Lithium atoms escape from a heated metal source and are captured and cooled by a magneto-optical trap (MOT).
The laser light for the MOT is then turned off, and the ultracold lithium atoms are trapped in the quadrupole magnetic field created by the permanent magnet rings. The CAVS is ready for sensing. After a hold time, the MOT laser light is turned on, and the remaining sensor atom number is determined from the atomic fluorescence. A new ensemble is then captured and cooled, and the process is repeated using a different hold time.} The distance between the bottom and top of this device is about 20~cm.}
        \label{fig:pcavs}
\end{figure}

In brief, the passage of a background-gas particle through the trapped cold ensemble, here $^{6}$Li, $^{7}$Li, or $^{87}$Rb alkali-metal gases, is detected by measuring the slow decrease of the number of sensor atoms in their trap. 
Given the immutable thermally averaged loss rate coefficients between the trapped ultracold atoms and the particles in the ambient background gas, the rate of collisions with trapped atoms, inferred from loss, is then related to the number density of the ambient gas. 
This positions cold-atom vacuum sensors (CAVS) as a novel paradigm for vacuum metrology. 
Figure \ref{fig:pcavs} shows a design of such a sensor {using a so-called quadrupole magnetic trap \cite{Metcalf1999}}, excluding the required laser systems and vacuum pumps, for operation at ambient temperatures.
As illustrated in Fig.~\ref{fig:CAVS_and_other_density-based_methods}, the sensor will operate at pressures below $10^{-4}$~Pa. 
{Design details can be found in Ref.~\cite{Ehinger2022}.
In fact, quadrupole magnetic traps are currently favored as they are simplest to implement.  
See also the design in Refs.~\cite{Booth2019,Shen_2020} from the University of British Columbia.
Measurements of loss rate coefficients quoted in this article have been performed using
sensor atoms in quadrupole magnetic traps.
}

After this short introduction,
we describe the history behind the cold atom vacuum sensor in more detail
in Section \ref{sec:cavsdetail}. In Sec.~\ref{sec:primary} we discuss why the cold atom vacuum sensor
is a primary standard for pressure.
Section
\ref{sec:determination_loss_rate} discusses several ways in which collisional rate coefficients have been obtained.
We describe theoretical limitations on the CAVS in Sec.~\ref{sec:discrepancies}
and constraints on implementations in Sec.~\ref{sec:implementations}.
Applications of the CAVS are described in Sec.~\ref{sec:application}.
We then describe our research plans for improving the CAVS in Sec.~\ref{sec:future}.
In Sec.~\ref{sec:alternativeideas} we discuss alternative ways to sense pressure in the ultra-high vacuum regime.
Finally, we conclude in Sec.~\ref{sec:conclusion}.

\section{The cold atom vacuum sensor}
\label{sec:cavsdetail}

\subsection{A brief history}
\label{sec:history}

Since the advent of laser cooling and trapping experiments, collisions with background gas particles in vacuum have been known to limit the lifetime of cold atomic ensembles \cite{Migdall1985,Bjorkholm1988}. The loss rate of ultracold sensor atoms $\Gamma_{\rm loss}$ from an idealized trap of vanishingly small depth due to background particles impinging on the ensemble is modeled by the differential equation
\be
{\rm d} n(t)/{\rm d}t= - \Gamma_{\rm loss} n(t)
\label{eq:lossrate}
\ee
for the number of ultracold atoms $n(t)$ as function of time $t$. In the limit of zero trap-depth, the loss rate satisfies
\be
\gloss = \gtot = \rn \langle v_{\rm rel}\sigma_{\rm tot}(E)\rangle \,,
\label{eq:rhosigmav}
\ee
where $\sigma_{\rm tot}(E)$ is the total collision cross section 
at relative collision energy $E$ between the cold atom and the background particle \cite{Mott1965,Sakurai1994,child1996molecular,Messiah1999}, $\gtot$ is the total collision rate,
and $v_{\rm rel}$ is the relative speed found from $E=\mu v_{\rm rel}^2/2$ with reduced mass $\mu=m_{\rm s}m_{\rm bg}/(m_{\rm s}+m_{\rm bg})$. Here, $m_{\rm s}$ and $m_{\rm bg}$ are the masses of the sensor atom and background particle, respectively. For later use, we also define collision
wavenumber $k_{\rm rel}$ with  $\hbar k_{\rm rel}=\mu v_{\rm rel}$ and
reduced Planck constant $\hbar$.
Brackets $\langle\cdots\rangle$ denote an average over Boltzmann distributions of three-dimensional momenta or velocities for the background particles at temperature $T$ and those
for the sensor atoms at temperature $T_{\rm s}$.  The walls of vacuum chambers are at ambient temperatures and $T_{\rm s}\lll T$.

For this roadmap, we can imagine a millimeter-sized cloud of $10^5$ ultracold atoms
at temperatures of several tens of microkelvin in a
shallow magnetic or optical trap with trap depths around or below $k\times 1$~mK.
In the UHV domain at room temperature, background number densities are $\rho\sim 10^6$ particles/cm$^3$ and assuming typical $\langle v_{\rm rel}\sigma_{\rm tot}(E)\rangle$, discussed later on in this article, decay times $1/\Gamma_{\rm loss}$ are of order 10~s to 100~s.

The authors of Refs.~\cite{PhysRevA.80.022712,PhysRevA.84.022708} showed that the loss rate depends on the depth of the sensor atom trap even for remarkably shallow traps with $U< \kb \times 100~\mu$K  and $\kb T_{\rm s}\lesssim U$.  In short, not every collision adds sufficient kinetic energy to a cold sensor atom to eject it from the trap.  In fact, based on the conservation of energy and momentum for elastic collisions and $\kb T_{\rm s}\ll U\ll \kb T$, these processes correspond to collisions for which the directions of the relative velocities before and after the collision only differ by less than a small, trap depth dependent, angle $\thetamin\approx 10$ mrad.  These encounters are so-called ``glancing angle'' collisions.

With the presence of glancing collisions, we realize that Eq.~(\ref{eq:rhosigmav}) needs to be replaced by a trap-depth-dependent loss rate 
\begin{equation}
     \gloss(U) = \gtot \left[ 1 - p(U) \right] \label{eq:glossU}
\end{equation}
where $p(U)$ is the probability that a collision imparts an energy $\Delta E \le U$ to the sensor atom.  Equivalently, we can define a trap-depth-dependent loss rate coefficient 
\begin{equation}
    \langle v_{\rm rel} \sloss(E,U)\rangle
    = \langle v_{\rm rel}\stot(E)\rangle \left[ 1 - p(U) \right]
    \,,
    \label{eq:rateU}
\end{equation}

Glancing-angle collisions require a quantum mechanical description even near room temperatures. This observation is easiest to see for a naive hard-sphere
model or potential with a closest approach at separation $a$. For example, from chapter 3 of Ref.~\cite{child1996molecular},
we read that the elastic differential cross section for this potential is highly peaked in the forward direction, favoring small changes in the directions of the relative velocities contrasting starkly with the isotropic prediction of a classical simulation.
In fact, the favored polar angular range is $\lesssim \theta_{\rm qd}\equiv 2/(k_{\rm rel}a)$, a ``quantum diffraction angle'', for $E\gg \hbar^2/2\mu a^2$. With $a\sim 1$~nm for our typical system parameters, we have $\thetamin \ll \theta_{\rm qd}$.

In 2011 the authors of Refs.~\cite{Booth2011, Madison2012} were the first to propose the use of ultra-cold atoms to realize a primary vacuum pressure standard. These authors argued that accurate measurements of the background number density can be made when the loss rate coefficient $\langle v_{\rm rel} \sigma_{\rm loss}(E,U)\rangle$ as a function of trap depth $U$ and temperature $T$ are known from either separate experiments or theoretical calculations.

Initial experiments to characterize ultracold atom loss and estimate the background gas number density were mostly focused on using magneto-optical traps (MOTs) with confinement provided by radiation pressure arising from the scattering of near-resonant laser light on one of the atom's electronic transitions \cite{PhysRevA.84.022708,PhysRevA.80.022712,PhysRevA.85.033420,yuan2013,Rev.Sci.Ins.093108.2015,pub.1105615573,Prentiss1988,Monroe1990, Steane1992,PhysRevA.64.023402, Matherson2007,xiang2018}. A quantitative connection between the loss rate to the background gas density was made uncertain by the large $\sim1$~K, but effectively unknown, depth of the MOT and, in some cases, the unknown populations of ground and excited electronic states of the sensor atoms in the MOTs. Moreover, the effects of the near-resonant MOT laser light on the collisions was and still is unknown.  Because the typical MOT depth of $\sim1$~K is much larger than the median collision energy imparted by room-temperature collisions, only about 1 in 5 background collisions induce trap loss.  A modern effort to control uncertainties in loss rate measurements in magneto-optical traps can be found in \cite{zhang2022,SUN2024113079}.

Parallel research demonstrated or proposed the use of atoms in shallow magnetic, invented in 1985 \cite{Migdall1985}, as an alternative without the complications of a MOT \cite{PhysRevA.80.022712,PhysRevA.84.022708,Booth2011, Madison2012,Julia2017,Julia2018,Eckel_2018,Booth2019} or optical dipole traps using focused off-resonant laser light \cite{Makhalov2016,Makhalov2017} to measure the number density of the background particles. 
For these experiments as well as for the CAVS, the atoms are first captured by and loaded into a MOT and then transferred to the shallower traps.
The ultracold atoms in these shallower traps are always in their electronic ground state and laser light, even when present, does not affect the collisions with the background gas particles.

Research prior to 2019 using trap loss rates to infer pressure was limited by the absence of accurate knowledge of the total collision rate coefficients and their dependence on trap depth. Consensus, however, already started to form that 
magnetic traps, relying on the magnetic moment of alkali-metal atoms and spatially varying magnetic fields, are the most convenient for pressure sensors.
These quadrupole \cite{Migdall1985,Bergeman1989} or Ioffe-Pritchard \cite{Pritchard1983} traps are shallow conservative potentials for atoms with controllable depths. Collisions occur in the absence of laser light and the determination of collision cross sections $\sigma_{\rm loss}(E,U)$ and thermalized rate coefficients $\langle v_{\rm rel} \sigma_{\rm loss}(E,U)\rangle$ is amenable to theoretical simulations.

In 2019, the first \emph{a priori} and first-principle numerical quantum scattering calculations of $\langle v_{\rm rel} \sigma_{\rm tot}(E)\rangle$ were performed. These simulations were performed  for $^6$Li and $^7$Li sensor atoms in their $^2$S electronic ground state and background H$_2$ molecules in thermally populated ro-vibrational levels of the X$^1\Sigma_g^+$ electronic ground state \cite{PhysRevA.99.042704,PhysRevA.105.039903}.  The authors found that the temperature dependence of the total rate coefficient 
for a two-temperature system
is solely a function of effective temperature
\begin{equation}
    T_{\rm eff} = \frac{m_{\rm s}}{m_{\rm s}+m_{\rm bg}} T
    +\frac{m_{\rm bg}}{m_{\rm s}+m_{\rm bg}} T_{\rm s} \,.
    \label{eq:Teff}
\end{equation}
This effective temperature is always less than $T$ and the contribution from the sensor atom temperature is negligible, especially given the typical experimental uncertainties in $T$.
The authors also showed
that inelastic rate coefficients for collision-induced transitions between rotational states of H$_2$ are much smaller than the uncertainties in the calculations of $\langle v_{\rm rel} \sigma_{\rm tot}(E)\rangle$. 

In the same year,  \cite{Madison2018,Booth2019} developed an analytical 
model for quantum diffractive collisions based on the isotropic long-range $-C_6/R^6$ interaction potential between the sensor atom and background particle. Here, $C_6$ is the van der Waals coefficient and $R$ is the separation between sensor atom and background particle.
They noted that, because of thermal averaging,  loss rate coefficients for room-temperature collisions are mostly insensitive to the interaction potential at small separations. In fact, the authors posited a universal quantum-diffractive collision (UQDC) model,
where the ratio $\langle v_{\rm rel} \sigma_{\rm loss}(E,U)\rangle/\langle v_{\rm rel} \sigma_{\rm tot}(E)\rangle$ is a polynomial in $U \times\langle v_{\rm rel} \sigma_{\rm tot}(E) \rangle/\langle v_{\rm rel}\rangle$ with coefficients that to good approximation are independent of the van der Waals coefficient, short-range potential, and temperature near ambient temperatures. 
In their papers, the model was used to estimate the total collision rate coefficients in the limit of zero trap depth for $^{87}$Rb sensor atoms with H$_2$, He, Ar, Xe, N$_2$, and CO$_2$ background gases.  

The universal model of Refs.~\cite{Madison2018,Booth2019} was validated for $^{87}$Rb+N$_2$ by a comparison to the pressure reported by an orifice-flow-calibrated ionization gauge \cite{Shen_2020,Shen_2021}. For $^{87}$Rb+$^{87}$Rb collisions \cite{PhysRevA.106.052812}, the model led to a van-der-Waals coefficient
that is consistent with the theoretical \cite{Derevianko2010} and
spectroscopic \cite{Roberts1998}  values in the literature.

Additional numerical quantum scattering calculations were reported  for Li+$^4$He  \cite{PhysRevA.101.012702,PhysRevA.105.029902}.  In 2021, the first direct comparison of two cold atom vacuum standards was reported \cite{Ehinger2022} while direct comparisons of a cold atom vacuum sensor with an orifice flow system were started \cite{BARKER2021100229,Barker2022}.
\emph{A priori} quantum mechanical scattering calculations of $\langle v_{\rm rel} \sigma_{\rm loss}(E,U)\rangle$ for $^{87}$Rb and $^7$Li colliding with multiple partners and including their $U$ dependence up to order $U^2$ were published in 2023~\cite{Klos:2023}. These authors also showed that for N$_2$ as well as for H$_2$
rotational (de)excitation processes have negligible contributions to their loss rate coefficients.

\cite{Shen_2023} observed the predicted breakdown of the universality method for light collision partners [for a discussion, see Refs.\cite{universality_revision,PhysRevA.110.063317,Guo2025}] using direct cross-species comparisons of $^{87}$Rb and $^{6}$Li sensor atom loss rates exposed to the same background vapor of H$_2$. These cross-species measurements were in good agreement with the \emph{a priori} quantum scattering calculations reported in Ref.~\cite{Klos:2023}. 

Finally,  \cite{Barker2023} reported measured thermalized collisional rate coefficients for $^7$Li and $^{87}$Rb colliding with room-temperature He, Ne, N$_2$, Ar, Kr, and Xe obtained by a combined cold atom sensor and an orifice flow vacuum standard, which can set a known pressure. Revised in 2025 to account for the non-zero temperature of the sensor-atom cloud, without changing the essential conclusions~\cite{eckel_effect_2025},  the measurements show consistency with the quantum-scattering calculations of Ref.~\cite{Klos:2023}. This consistency provides the required validation that a quantum-based cold-atom vacuum standard serves as both a sensor and primary standard for vacuum.

\subsection{Summary of measured and calculated loss rate coefficients for $^7$Li- and $^{87}$Rb-based CAVSs}

\begin{table}
\begin{tabular}{cD{.}{.}{2.6}D{.}{.}{2.6}}
\hline\hline
Background   & \multicolumn{2}{c}{$\left<v_{\rm rel}\sigma_{\rm{loss}} 
 (E,U)\right>$  for $^7$Li} \\
\multicolumn{1}{c}{gas} &  \multicolumn{2}{c}{($10^{-9}$cm$^3$/s)} \\
   & \multicolumn{1}{c}{FQMC} & \multicolumn{1}{c}{OFS
  }  \\
\hline
  H$_2$  & 3.18(6) &\\
 $^4$He & 1.66(4) & 1.72(3) \\
 Ne & 1.6(1) & 1.63(2) \\
 N$_2$ & 2.66(2) & 2.72(3)  \\
 Ar & 2.36(1) & 2.38(2) \\
 Kr & 2.166(7) & 2.18(3) \\
 Xe & 2.27(2) & 2.22(3) \\
\hline
\end{tabular}
\caption{
Loss rate coefficients $\left<v_{\rm rel}\sigma_{\rm loss}(E,U)\right>$ of $^7$Li due to collisions with various species in the background gas at trap depth $U=k\times0.95(14)$~mK as determined by first-principle quantum scattering calculations (FQMC)~\cite{Klos:2023} and measured with a CAVS using an orifice flow standard (OFS) to set known pressures~\cite{Barker2023}. The temperature of the background particles is 301.7(3.3)~K.  The numbers in parenthesis are combined systematic and statistical $k=1$ one-standard deviation uncertainties.}
\label{tab:Li_summary} 
\end{table}

Tables \ref{tab:Li_summary} and \ref{tab:Rb_summary} summarize the most accurate measurements and calculations to date of the loss rate coefficients $\langle v_{\rm rel}\sigma_{\rm loss}(E,U) \rangle$ for $^7$Li and $\langle v_{\rm rel}\sigma_{\rm tot}(E) \rangle$ for $^{87}$Rb, respectively, induced by collisions with  several background species. 
The measurements for $^7$Li agree within the quantum scattering calculations within 2~\%, with the exception of $^7$Li-$^4$He, where the discrepancy is slightly larger, 3.6~\%.
Remarkably, despite several markedly different techniques being brought to bear, agreement on the values of $\langle v_{\rm rel}\sigma_{\rm tot}(E) \rangle$ for $^{87}$Rb-X collisions is better than 17~\%.
The UQDC value  for $^{87}$Rb+H$_2$ is an outlier because of the expected breakdown in the universality hypothesis for light mass species described earlier. 
We observe disagreement between the various techniques at the 3-$\sigma$ level or larger, suggesting that in one or more of the measurement techniques, uncertainties are underestimated or there is some unknown systematic effect that remains unaccounted for.
The discrepancies require further investigation.

\begin{table}
\begin{tabular}{cD{.}{.}{2.6}D{.}{.}{1.4}D{.}{.}{2.6}D{.}{.}{1.4}}
\hline\hline
Background  & \multicolumn{4}{c}{$\svtot$ ($10^{-9}$cm$^3$/s) for $^{87}$Rb}  \\
gas & \multicolumn{1}{c}{UQDC} & \multicolumn{1}{c}{Ratiometric} & \multicolumn{1}{c}{FQMC } &\multicolumn{1}{c}{OFS
  }  \\
\hline 
 H$_2$ &  5.12(15) & 3.8(2) &  3.9(1) & \\
 $^4$He &  2.41(14) &   & 2.37(3) & 2.34(6) \\
 Ne & &  & 2.0(2) & 2.21(5) \\
 N$_2$ & 3.14(5) &   & 3.45(6) &3.56(8) \\
 Ar &  2.79(5) &   & 3.035(7) & 3.31(5) \\
 CO$_2$ & 2.84(6) &   & &  \\
 Kr &   &  & 2.79(1) & 2.79(4) \\
 Xe &  2.75(4) &  & 2.88(1) & 2.95(7) \\
\hline
\end{tabular}
\caption{
Loss rate coefficients at zero trap depth $\langle v \sigma_{\rm loss}(E,U=0) \rangle=\langle v \sigma_{\rm tot}(E) \rangle$
and $T = 294(1)$~K for $^{87}$Rb induced by collisions with various species in a background gas as determined by the universal scaling of quantum diffractive collisions (UQDC)~\cite{Shen_2021}, ratiometric techniques~\cite{Shen_2023}, first-principle quantum scattering calculations (FQMC)~\cite{Klos:2023}, and comparison with an orifice flow standard (OFS)~\cite{Barker2023}. The numbers in parentheses are combined systematic and statistical one-standard deviation $k=1$ uncertainties. }
\label{tab:Rb_summary}
\end{table}

As of the writing of this roadmap, an independent study is underway to measure the total collision rate coefficients for $^{87}$Rb sensor atoms and a variety of gases using an orifice flow system at Physikalisch-Technische Bundesanstalt (PTB) in Berlin. 
Preliminary results of Ref.~\cite{Waldock2024} are in good agreement with the experimental values reported in Ref.~\cite{Barker2023}, further validating the CAVS as a primary standard.

\section{Criteria for a primary standard} \label{sec:primary}

We have already implied that the cold atom vacuum sensor can serve as a primary standard for pressure.
To see that this is true, it is useful to refer to the definition of a primary standard in the International Vocabulary of Metrology~\cite{VIM}. 
First, the measurement procedure for the realized quantity must be without relation to a measurement standard for the quantity of the same kind.
For a CAVS, this means that there must be no relation to another measurement standard for number density or pressure. 
In practice, this requirement is met with first-principle quantum mechanical calculations of $\svtot$ and $\svloss$ as functions of background gas temperature $T$ and depth $U$ of the traps that hold the sensor atoms.
The universal model for the trap depth dependence of $\svloss/\svtot$ can play a similar role.

Second, the uncertainty of the measurement method must be quantified.
Practically, uncertainty quantification requires that the physics \textit{and} technology of the measurement method is completely understood so that an operational equation can be derived.
The operational equations for a CAVS are Eqs.~(\ref{eq:lossrate}) and (\ref{eq:rhosigmav}), or their generalizations Eqs.~(\ref{eq:glossU}) and (\ref{eq:rateU}), which
account for the depth $U$ of the traps that hold the sensor atoms.
Temperature $T$ and depth $U$ in these equations can be set or measured with quantifiable uncertainties.

Third, the measurement standard must be accepted as a primary standard by the international metrology community, which typically requires independent validations of the measurement standard.
Such independent validations can be accomplished by comparisons with other measurement standards of the same or similar kind or, in the ideal case, with measurement standards applying a different measurement method. 
This condition can be met as there exists a pressure region where the cold-atom vacuum sensor and the orifice flow standard for pressure operate at similar 1\,\% relative accuracy.
This pressure region is around $10^{-6}$~Pa.

Additional requirements for a measurement standard include the possibility of a wide dissemination of the measured quantity, for example, by calibration of a measurement instrument.
It is not useful when the quantity can only be realized in a specialized laboratory.
The results obtained by the measurement standard must be reproducible at different times, by different operators and at different locations.
This implies that the measurement standard must be robust and not depend on specialized environmental conditions or require specialized experience for the operators.
With the successful build and operation of the portable CAVS shown in Fig.~\ref{fig:pcavs} parts of these requirements have been met.

\section{Determination of the rate coefficients}
\label{sec:determination_loss_rate}

The cold atom vacuum sensor relies on accurate knowledge of the total collision rate coefficient $\svtot$ and loss rate coefficient $\svloss$.
A theoretical, an experimental, a hybrid, and a ratiometric method for determining these rate coefficients exist.
The theoretical method is formally exact and corresponds to full quantum mechanical scattering calculations,
where first the relevant potential energy surfaces for the (relativistic) electron motion as function of frozen nuclear positions are computed and then the atomic or nuclear motion is solved as coupled Schr\"odinger equations using these potential energy surfaces to determine rate coefficients as function of temperature. 
This approach has, for example, been taken in  Refs.~\cite{Klos:2023,PhysRevA.105.029902,PhysRevA.99.042704,PhysRevA.105.039903,PhysRevA.101.012702}.
Its accuracy is only limited by our ability to compute potential energy surfaces, to numerically solve for the solutions of coupled Schr\"odinger equations,
and to account for non-adiabatic recoil corrections coupling the electron and nuclear motion.
   
The experimental method is to generate a known  pressure for each relevant background gas, by accurate conventional methods, {\it e.g.} an orifice flow standard~\cite{Jousten1999, Barker2022}.  This pressure must be low enough such that the
cold-atom vacuum sensor remains operational. The sensor-atom loss rate in the limit of zero trap depth can then be used to determine $\svtot$ by combining Eqs.~(\ref{eq:idealgaslaw}) and (\ref{eq:rhosigmav}) \cite{Barker2023}. 
The accuracy of this approach is determined by the accuracy of the conventional method
and the uncertainty in the measured loss rates.
In addition, we note that limited information regarding the temperature dependence
can be extracted experimentally as changing the temperature of the vacuum chamber over
more than 10~$^\circ$C is impractical.

The hybrid theoretical and experimental option consists of combining 
measurements of $\Gamma_{\rm loss}(U)$ as function of trap depth $U$ {\it without} knowing the background pressure or number density with the expected analytical behavior of this rate from the universal diffractive collision model. As mentioned in Sec.~\ref{sec:history}, the analytical behavior relies on the observation that, to good approximation near room temperature, the {\it ratio} of the loss and total
rate coefficients and thus also $\Gamma_{\rm loss}(U)/\Gamma_{\rm loss}(U\to0)$ as functions of $U$ are insensitive to the details of the electronic potential energy surfaces at small atom-atom or atom-molecule separations \cite{Booth2019,Shen_2020}. 
The accuracy of this approach is limited by the uncertainties in $\Gamma_{\rm loss}(U)$ and the approximations inherent in the universal diffractive collision model.

A final means to measure rate coefficients is to co-trap ensembles of two species of sensor atoms, say Li and Rb, so that both sensor species are exposed to the same background gas number density at the same background temperature. After measuring the rates $\Gamma_{\rm loss}$ for both species, the ratio of these rates is equal to a ratio of the loss rate coefficients independent of the background pressure.

We describe the four techniques for measuring or calculating $\svtot$ and $\svloss$ in the following subsections.

\subsection{Method I: First-principle quantum calculations of rate coefficients}

In this section, we describe the minimal framework for determining $\svtot$ and $\svloss$ from first-principle quantum simulations.
We can omit the internal structure of the collision partners.
These include the hyperfine splittings of the ground-state alkali-metal sensor atoms as well as the rotational and vibrational states and fine- and hyperfine-structure of atoms and molecules in the vacuum.
Hyperfine splittings are on the order of $\kb\times 0.1$~K, which is much smaller than the background gas temperatures.
Inert noble-gas atoms, often deliberately introduced into a vacuum chamber, have no internal structure.
Inert molecules, like the tightly-bound H$_2$ and N$_2$ with their $^1\Sigma_g^+$ electronic ground state \cite{PhysRevA.99.042704,Klos:2023}, have negligible fine and hyperfine structure and their electronic interaction potential with the sensor atom is to good approximation spherically symmetric, leading to weak couplings between rotational and vibrational states.
With these assumptions, we only need to describe elastic, momentum-changing collisions. 

\begin{figure}
   \includegraphics[scale=0.35, trim= 0 25 0 60,clip]{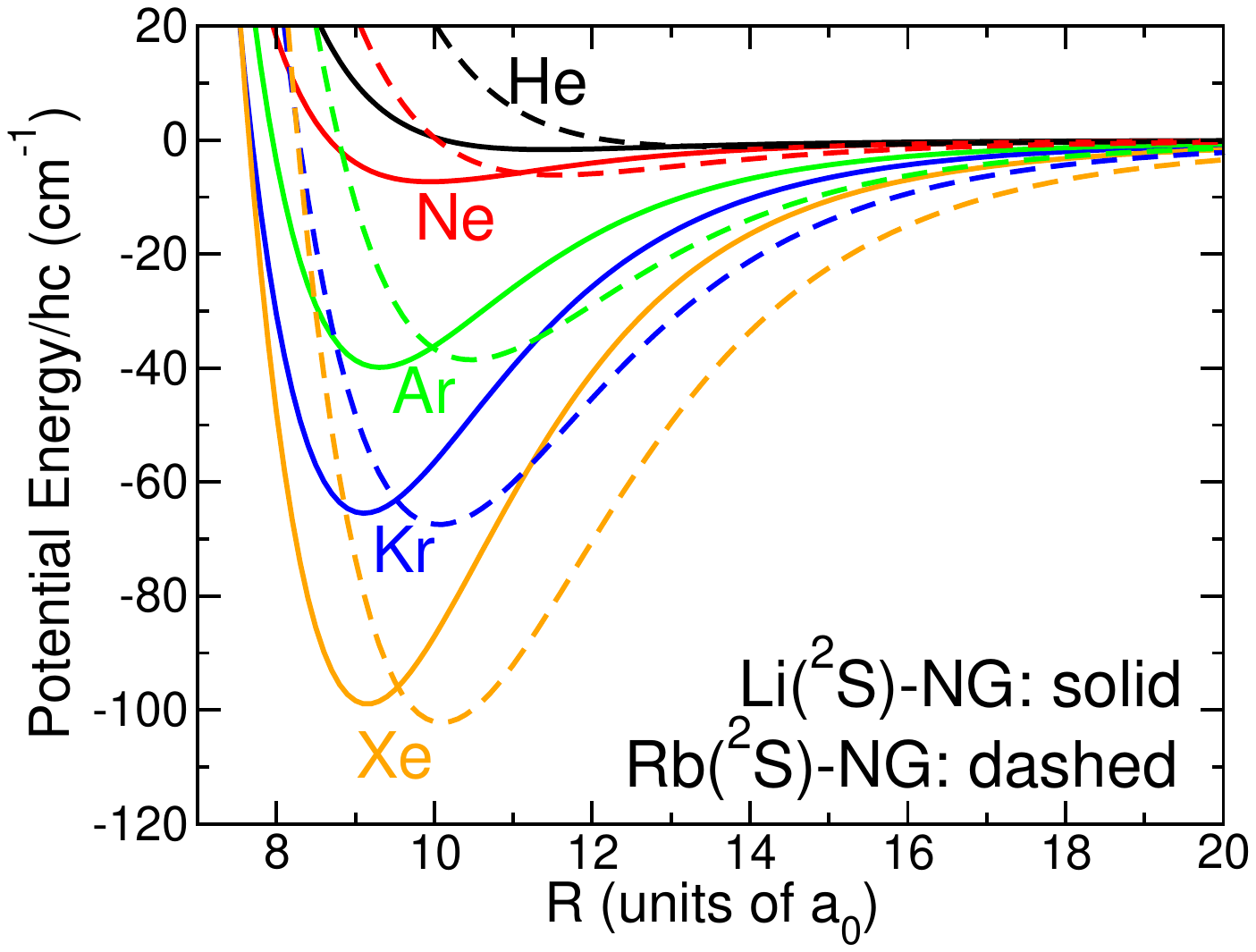} 
        \caption{The most-accurate calculated electronic potentials  as functions of inter-particle separation $R$ for Li($^2$S)-NG (solid curves) and  Rb($^2$S)-NG (dashed curves)  used in first-principle quantum calculations of loss-rate coefficients, where NG stands for Noble Gas. Taken from Ref.~\cite{Klos:2023}.
        } \label{fig:PESs}
\end{figure}

We thus consider an elastic collision between a stationary, trapped sensor atom and a background gas particle whose velocity is selected from a three-dimensional Maxwell-Boltzmann distribution at ambient temperature. The collision is analyzed in the center of mass frame and scattering occurs from a spherically symmetric or isotropic interaction potential $V(R)$, where $R$ is the separation between the center of masses of the sensor atom and background particle. 
For separations $R\gtrsim 20a_0$, $V(R)$ approaches the attractive dispersion potential $-C_6/R^6$ with positive van der Waals coefficient $C_6$. Here, $1a_0=0.052\,9177$ nm is the Bohr radius. Figure~\ref{fig:PESs} shows the short-range potentials for our sensor atoms interacting with the spin-less noble gases. The potentials are relatively shallow with depths less than $hc\times 100$ cm$^{-1}$, where $h$ is the Planck constant and $c$ is the speed of light in vacuum.
The van der Waals coefficients for collisions with noble gases are taken from Ref.~\cite{Derevianko2010}, while those for H$_2$ and N$_2$ can be found in Ref.~\cite{PhysRevA.99.042704,Klos:2023}.

Elastic scattering conserves energy and momentum and, in the center of mass frame, only the direction of the relative momentum of the collision partners is altered, rotating by 
collision or polar angle $\theta\in[0,\pi]$ and azimuthal angle $\varphi\in[0,2\pi]$ \cite{child1996molecular}. Moreover, it is straightforward to show that for an initial relative speed $v_{\rm rel}$, the kinetic energy transferred to a stationary sensor atom in a collision is given by
\be
\Delta E = \frac{\mu^2 v_{\rm rel}^2}{\mt}\left(1-\cos\theta\right)\,,
\label{eq:DeltaEtrapped}
\ee
independent of azimuthal angle.
For a sensor atom at rest, we also have $v_{\rm rel}=v_{\rm{bg}}$,
where $v_{\rm{bg}}$ is the initial speed of the background particle.

The sensor atom is ejected from the trap when the energy transfer $\Delta E$ is greater than the trap depth $U$. Rearranging Eq.~(\ref{eq:DeltaEtrapped}) and equating $\Delta E = U$, we derive that the  minimum scattering angle $\thetamin$ required for trap loss is given by
\be
\cos\thetamin =  1 - \frac{\mt U}{\mu^2 v_{\rm rel}^2}.
\label{eq:thetamin}
\ee

Observables of a scattering process, such as cross sections and rate coefficients,
are found from wavefunctions $\psi({\bf r})$ that are numerical solutions of the three-dimensional Schr\"odinger equation for motion in isotropic potential $V(R)$. Vector ${\bf r}=(x,y,z)=(R,\theta,\varphi)$ is the three-dimensional relative coordinate, where the two equalities specify ${\bf r}$ in Cartesian and spherical coordinates, respectively.
In the limit $R\to\infty$, this wavefunction for relative kinetic energy $E$ or relative wavenumber $k_{\rm rel}$ is the superposition of an incident plane wave along the $z$ direction and a scattered spherical wave, $\psi({\bf r}) \rightarrow e^{ik_{\rm rel}z} + f(k_{\rm rel},\theta) e^{ik_{\rm rel}R}/R$, where length $f(k_{\rm rel},\theta)$ describes the  wave amplitude into angle $\theta$ relative to the $z$ direction. For spherically symmetric potentials, $f(k_{\rm rel},\theta)$ does not depend on azimuthal angle $\varphi$.

The scattering amplitude $f(k_{\rm rel},\theta)$ can be expressed as a sum of Legendre polynomials $P_L\left(x\right)$ weighted by dimensionless, complex-valued transition matrix elements $f_L(k_{\rm rel})$, corresponding to contributions from ``partial waves" with angular momentum $\hbar L$. That is, we have
\be
f(k_{\rm rel},\theta) = \frac{1}{k_{\rm rel}}\sum_{L=0}^\infty \left(2L+1\right) f_L(k_{\rm rel}) P_L\left(\cos\theta\right).
\label{eq:fktheta}
\ee
The total elastic cross section for a scattering event with wavevector $k_{\rm rel}$ is then
\bea
\sigma_{\rm{tot}}\left(E\right) 
&=& \int_0^\pi \sin\theta  {\rm d}\theta \int_0^{2\pi} {\rm d}\varphi \left| f(k_{\rm rel}, \theta)\right|^2 \,,
\label{eq:sigmatotk}
\eea
while the thermally averaged total collision rate coefficient is 
\bea
\nonumber
\svtot 
& = & \left< 2\pi  v_{\rm rel} \int_{0}^{\pi} \sin\theta {\rm d}\theta \left|f\left(k_{\rm rel}, \theta\right)\right|^2 \right>\,.
\label{eq:svtot}
\eea
The brackets $\langle\cdots\rangle$ imply an average over the Maxwell-Boltzmann distribution 
with effective temperature $T_{\rm eff}$ in Eq.~(\ref{eq:Teff}) with $T_{\rm s}\to0$. 
Within our uncertainty budgets, this simplification  is justified for a CAVS because the temperature of the sensor atoms is about $10^6$ times smaller than the temperature of the background gas particles.

Trap loss of sensor atoms is guaranteed for $\theta \in [\thetamin,\pi]$, leading to the thermally averaged trap loss rate coefficient
\be
\svloss = \left< 2\pi  v_{\rm rel} \int_{\thetamin}^{\pi} \sin\theta {\rm d}\theta \left| f\left( k_{\rm rel}, \theta \right) \right|^2 \right>\,,
\label{eq:svloss}
\ee
where $\thetamin$ depends on sensor-atom trap depth $U$ and relative collision speed $v_{\rm rel}$. 
We rewrite Eq.~(\ref{eq:svloss}) as
\bea
\svloss &=& 
 \svtot \left[ 1 - p(U) \right]
\label{eq:svlosspU}
\eea
with
\be
p(U) = \frac{\displaystyle\left< 2\pi  v_{\rm rel}\int_{0}^{\thetamin} \sin\theta {\rm d}\theta \left|f\left(k_{\rm rel}, \theta\right)\right|^2 \right>}{\svtot}\,,
\label{eq:pU}
\ee
the cumulative probability that a collision imparts energy $\Delta E \le U$ to the sensor atom.
Clearly, $p(U)\to 0$ for $U\to 0$.

For several systems, the authors of \cite{PhysRevA.105.029902,PhysRevA.99.042704,PhysRevA.105.039903,PhysRevA.101.012702,Klos:2023} computed $f_L(k_{\rm rel})$ for collision energies up to six times $\kb T_{\rm eff}$ and partial waves $L$ up to a few hundred using numerical solutions of the Schr\"odinger equation for each $L$ and constructed $\svtot$ as well as the linear and quadratic dependence of $p(U)$
as function of $U$. 
Some of their results can be found in Tables \ref{tab:Li_summary} and \ref{tab:Rb_summary}. The authors relied on new evaluations of the isotropic electronic potentials $V(R)$ including a determination of their $R$-dependent uncertainties. These uncertainties led to the few percent uncertainties in the observable rate coefficients.
They also realized that for H$_2$ and N$_2$ the contribution to the total rate coefficients from the anisotropic components of the electronic
potential is smaller than the uncertainties in the published rate coefficients.

\subsection{Method II: The universality hypothesis for room-temperature collisions}
\label{sec:universality-method}

Evidence that $\svtot$ and $\svloss$ are insensitive to the shape of the interaction potential $V(R)$ for $R\lesssim 20a_0$ was reported in Refs.~\cite{Madison2018,Booth2019, Shen_2020, Shen_2021} and further explored in Refs.~\cite{universality_revision,PhysRevA.110.063317,Guo2025}. One reason for this ``universal'' behavior is related to the fact that small-angle scattering and small deflections at ambient velocities are predominantly due to the long-range van der Waals tail of the interaction potential \cite{child1996molecular}. Based on this and the optical theorem
\begin{equation}
    \sigma_{\rm tot}(E)= \frac{4\pi}{k_{\rm rel}} {\rm Im}f(k_{\rm rel},\theta=0)\,,
    \label{eq:opticaltheorem}
\end{equation}
one might naively assume that the total collision rate coefficient $\svtot$ should also be predominantly determined by the van der Waals tail of the potential.  However, the total cross section is completely dominated by the shape of the interaction for $R\lesssim 20a_0$ for sufficiently energetic  collisions.  For room-temperature collisions, the influence is still significant; however, because of the velocity averaging inherent in the rate coefficient, this influence is averaged away.  In short, velocity averaging is the origin of universality.  Hence, $\svtot$  for two interaction potentials with the same long-range behavior are nearly the same.  This phenomenon has the fortunate consequence of significantly reducing errors determining $\svtot$ arising from systematic errors in the \emph{ab initio} calculated potential.

Equally important for the analysis of loss rate coefficients for sensor atoms held in a trap with a finite depth is the observation that, to a good approximation, the cumulative probability $p(U)$ in Eq.~(\ref{eq:pU}) can be described by a (finite) power series in $U\svtot$ \cite{Booth2019}. That is, we have
\be
p(U) = \sum_{j=1}^N \beta_j \left(\frac{U}{\ud}\right)^j\,,
  \label{eq:pqdu}
\ee
where the characteristic energy scale $U_{\rm d}$ is the median kinetic energy imparted to an initially stationary sensor atom. That is,
\be
\ud = \frac{4\pi \hbar^2}{\mt \bar{\sigma}} \,,
\label{eq:Udval}
\ee
where $\bar{\sigma} = \svtot/\vp$ is the averaged cross section, and $\vp=\sqrt{\pi}\langle v_{\rm rel}\rangle/2=\sqrt{2\kb T/m_{\rm bg}}$ is the speed at the peak of the Maxwell-Boltzmann distribution for the background particles.
The energy scale $U_{\rm d}$ depends on the temperature of the background gas.
As follows from the data in Tables \ref{tab:Li_summary} and \ref{tab:Rb_summary}, $U_{\rm d}\approx\kb \times 10 \; \mathrm{mK}$ for $^{87}$Rb sensor atoms and $U_{\rm d}\approx\kb\times 100 \; \mathrm{mK}$ $^7$Li sensor atoms at ambient temperatures.
For both sensor atoms the experimental trap depths are much smaller than $U_{\rm d}$.

The dimensionless coefficients $\beta_j$ are assumed to be common to all collision partners and independent of $T$. They  are listed in Table~\ref{tab:betajs} for $N=6$ and found from numerical experiments by Ref.~\cite{Booth2019}. The corresponding $p(U)$ is valid for $U/U_{\rm{d}} \le 2$.
These coefficients were obtained by a polynomial fit to the right hand side of Eq.~(\ref{eq:pU}) computed by quantum scattering calculations for Li and Rb sensor atoms colliding with three of the noble gas atoms using  Lennard-Jones potentials 
with  $C_6$ values from \cite{Derevianko2010} and $C_{12}$ coefficients such that the Lennard-Jones potential approximately reproduces the published depths of the physical potentials $V(R)$.
The uncertainties of the $\beta_j$s were found from the distribution of their values for the six studied systems.

\begin{table}
\begin{tabular}{c@{\ }r@{.}l}
\hline
\hline
 Coefficient &  \multicolumn{2}{c}{Value}   \\
\hline
$\beta_1$ & 0 & 673(7) \\
$\beta_2$ & $-$0 & 477(3)\\
$\beta_3$ & 0 & 228(6)\\
$\beta_4$ & $-$0 & 0703(42)\\
$\beta_5$ & 0 & 0123(14)\\
$\beta_6$ & $-$0 & 0009(2)\\
\hline
\end{tabular}
\caption{Universal coefficients $\beta_j$ with standard uncertainties in parentheses as defined in Eq.~(\ref{eq:pqdu}). Taken from Ref.~\cite{Booth2019}.
}
\label{tab:betajs}
\end{table}

A consequence of the universality of $\svloss$ is that  measurements of trap loss rate $\Gamma_{\rm loss}(U)$ as a function of trap depth  at the same background number density and temperature can be fit to Eq.~(\ref{eq:pqdu}) with $\svtot$ as the only adjustable parameter.  
This universality method was used in Refs.~\cite{Booth2019,Shen_2020,Shen_2021} to obtain  total collision rate coefficients.  In Table~\ref{tab:Rb_summary} these results are compared with those obtained from first-principle quantum mechanical calculations and from loss rate measurements at known background gas number densities using an orifice flow standard.  
The agreement is rather remarkable, except for the light H$_2$.

Still, the observed discrepancies motivated a closer examination of the universality hypothesis \cite{universality_revision,PhysRevA.110.063317}.  
The outcome was confirmation that the analyses are sound as long as the velocity distribution of the background particles above a critical velocity does not have significant weight but also that additional attractive dispersion interactions, namely $-C_8/R^8$ and $-C_{10}/R^{10}$, do contribute to the total collision rate coefficient. These contributions limit the universality of the coefficients $\beta_j$.  

We finish these two subsections by comparing  rate coefficients from the first-principle quantum calculations with two approximate semiclassical models based on small-angle scattering from a long-range van der Waals potential and one where all three dispersion interactions are included. 
The first of these semi-classical models starts from an approximate representation of the matrix element $f_L(k_{\rm rel})$. It is solely a function of $(k_{\rm rel}x_6)^4/(L+1/2)^5$,
where $x_6=\sqrt[4]{2\mu C_6/\hbar^2}$ is the van der Waals length.
A brief derivation, found for example in Ref.~\cite{child1996molecular}, using the Euler-Maclaurin formula applied to the sum over partial waves $L$ in Eq.~(\ref{eq:fktheta}) then shows that the  thermalized total elastic rate coefficient  has a closed form and is given by
\be
\svtotCsix = q_{\rm C6}   \left(\frac{C_6}{\hbar \vp}\right)^{2/5} \vp 
\label{eq:svtot6JB}
\ee 
with $q_{\rm C6}=2^{4/5}3^{2/5}\pi^{9/10}\sin(3\pi/10) \Gamma(3/5)\Gamma(9/5)\approx8.495$ and $\Gamma(z)$ is the Gamma function. For later use, it is worth noting that $\svtotCsix \propto (\kb T/m_{\rm bg})^{3/10}$, a slowly increasing function of the temperature of the vacuum chamber and independent of the sensor atom mass.

The second semiclassical model starts from an approximate representation for $f_L(k_{\rm rel})$ that also includes contributions from dispersion potentials $-C_8/R^8$ and $-C_{10}/R^{10}$ with coefficients where available from Ref.~\cite{Jiang2015}.
In this case, as observed in Ref.~\cite{universality_revision} the sum over $L$ in the imaginary part of $f(E,\theta=0)$ as well as the thermalization with a Boltzmann distribution must be computed numerically. We denote this thermalized elastic rate coefficient by $\svtot_{\rm{disp}}$.

\begin{figure}
\includegraphics[width=\linewidth]{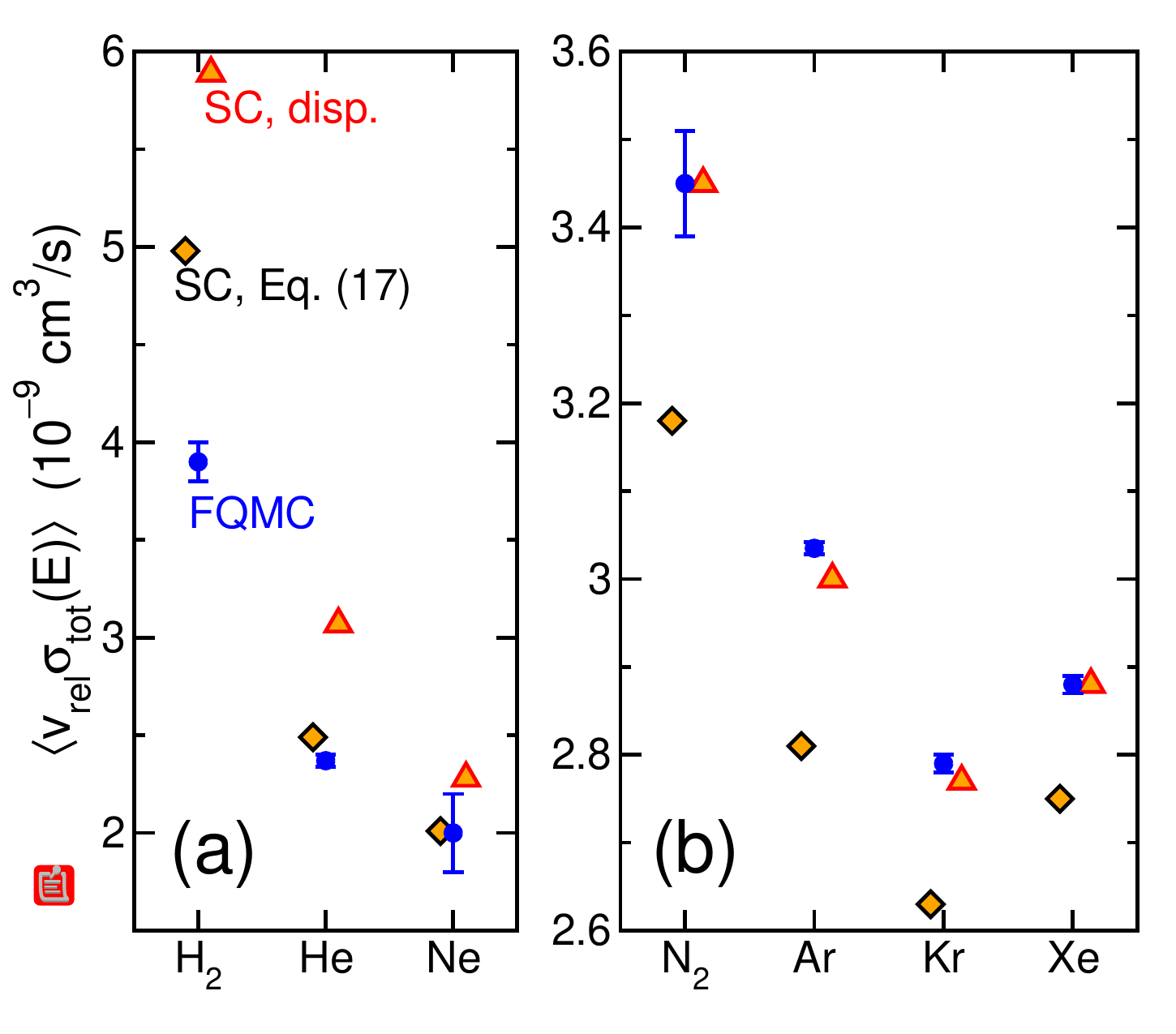} 
\caption{
A comparison of $\svtot$ at $T = 294(1)$~K from the first-principle quantum calculations (FQMC, 
{solid blue circles with error bars}) with standard uncertainties from Ref.~\cite{Klos:2023} with two semiclassical models for systems $^{87}$Rb-$X$ with $X={\rm H}_2$, He, Ne in panel (a) and $X={\rm N}_2$, Ar, Kr, and Xe in panel (b). Black diamonds and red triangles show data for the semiclassical model in Eq.~(\ref{eq:svtot6JB}) and the semiclassical model including effects of multiple dispersion coefficients \cite{universality_revision,PhysRevA.110.063317}, respectively.  Note the difference in the vertical scales for the two panels.}
\label{fig:svtotexpC6}
\end{figure}

\begin{figure}
\includegraphics[width=\linewidth]{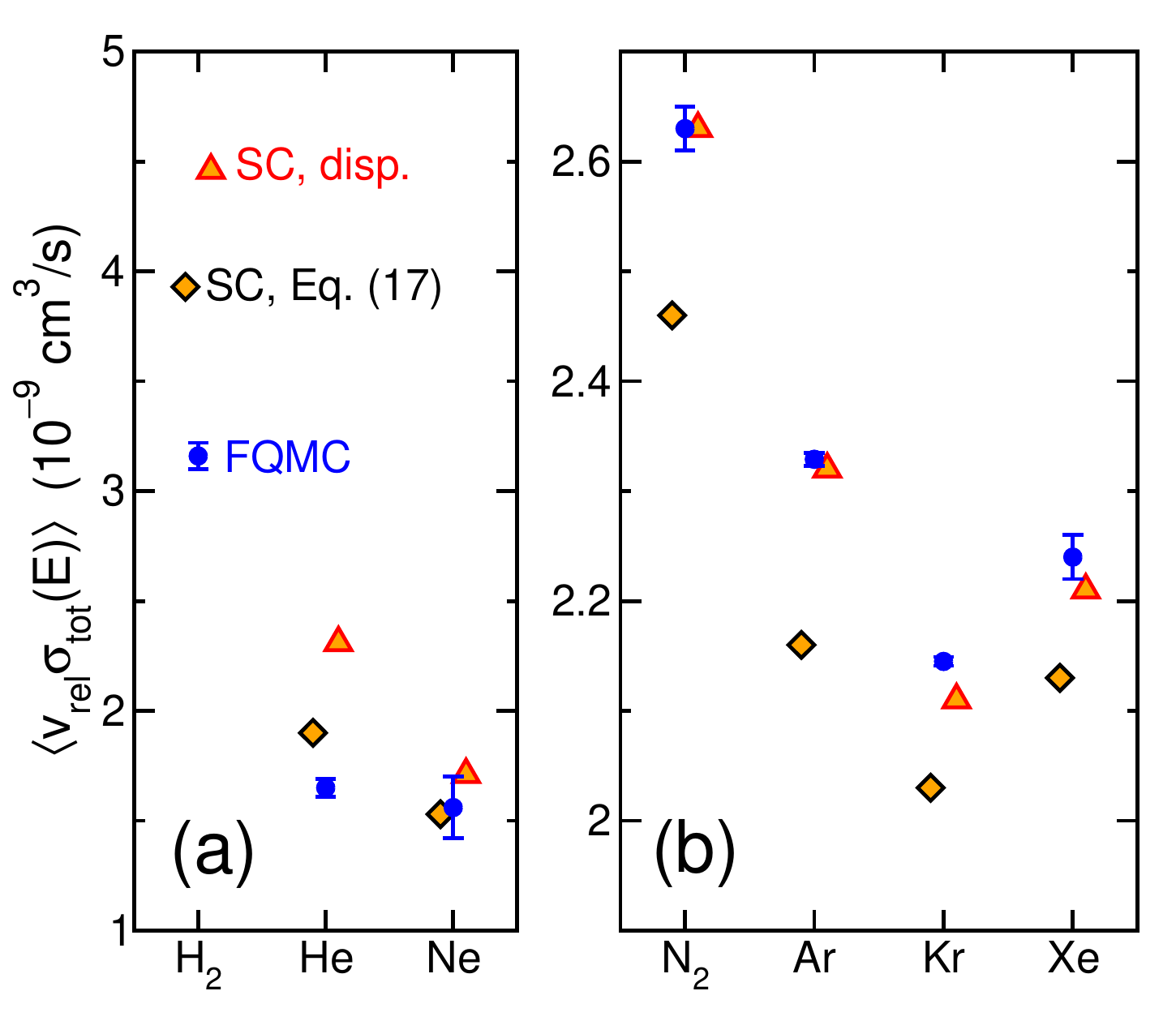} 
\caption{
{
A comparison of $\svtot$ at $T = 294(1)$~K from the first-principle quantum calculations  with standard uncertainties  with two semiclassical models for systems $^{7}$Li-$X$ with $X={\rm H}_2$, He, Ne in panel (a) and $X={\rm N}_2$, Ar, Kr, and Xe in panel (b). 
Marker definitions and data origin are the same as those in Fig.~\ref{fig:svtotexpC6}.
Note the difference in the vertical scales for the two panels.
}
}
\label{fig:svtotexpC6Li}
\end{figure}

{
Figures \ref{fig:svtotexpC6} and \ref{fig:svtotexpC6Li} compare thermalized $\svtot$ from the first-principle quantum calculations for $^{87}$Rb and $^7$Li sensor atoms, respectively, with the two semiclassical models.}
Generally, the agreement between the first-principle quantum calculations and the two semiclassical models improves with increasing reduced mass of the system.
In fact, the rate coefficients based on the second semiclassical model with its better modeling of the attractive long-range interactions are noticeably in better agreement with the first-principle  calculations for the heavier systems N$_2$, Ar, Kr, and Xe.
On the other hand, for the light systems H$_2$ and He, the second model is worse,
indicating that these $\svtot$  are significantly impacted by the core repulsion of the potentials. For a neon background gas, both semiclassical models agree with the first-principle calculation as the quantum calculations for Ne have a relatively large uncertainty.

\subsection{Method III: Measurement by comparison with other vacuum metrology standards}

Measurements of loss rate coefficients $\svloss$ can be accomplished by comparing a CAVS to other vacuum metrology standards,  for example, an orifice flow standard~\cite{Tilford1988, Jousten1999}.
Orifice flow standards (OFS) add a known number density of a known gas at a known background temperature into a vacuum chamber initially prepared at base pressure $p_{\rm base}$ according to 
\begin{equation}
   \Delta \rn = \dot{N}/S\,,
\end{equation}
where $\dot{N}$ is the flow of gas into the chamber measured in particle number per second and $S$ is the pumping speed out of the chamber in units of L\,s$^{-1}$.
The flow $\dot{N}$ is both generated and measured using a constant-pressure flowmeter~\cite{Jousten1993, Jousten2002, Eckel2022}.
The pumping speed is set by restricting the flow out to a single aperture with a known conductance $C_0$. Hence, $S=C_0$.
For an OFS, the relative uncertainty in the added $\Delta\rn$ is $<0.5$\,\%.
The comparison with a CAVS can succeed, {\it i.e.} have sufficient accuracy, when
the value and uncertainty of the number density $p_{\rm base}/\kb T$ at base pressure are significantly smaller than the added number density. 
We can either expect $u(p_{\rm base})\ll p_{\rm base}$ for a reproducible but unknown base pressure or $u(p_{\rm base})\approx p_{\rm base}$ for a fluctuating but small base pressure.

The comparison  proceeds by injecting a known amount of background gas and measuring the loss rate $\Gamma_{\rm loss}(U)$ at known, measured depth $U$. In fact, we determine
$\Gamma_{\rm loss}(U)$ for several $\Delta\rn$ and $U$
and fit $\Gamma_{\rm loss}(U)$ to a linear function in $\Delta\rn$, with an offset to account for the contribution from residual gases at base pressure, and a low-order polynomial in $U$. The fit gives us  $\svloss$.
This approach was adopted in Ref.~\cite{Barker2023}, and enabled a comparison with the theoretical values for $\svloss$ or those derived by other experimental means.
{
In fact, a study of Tables~\ref{tab:Li_summary} and \ref{tab:Rb_summary} shows that
results from first-principle quantum calculations and experimental data obtained with the orifice flow standard are consistent except for the data for  $^{87}$Rb sensor atoms with background Ar atoms. These deviate by almost five sigma and should be revisited both experimentally and theoretically. }

The fitting procedure is also useful for gases for which either a theoretical prediction is unavailable or the application of the universality hypothesis is questionable.
In these cases, comparison with the derived value of $\svloss$, if subsequently used for a pressure measurement, would not constitute a primary measurement. 
However, as shown in Ref.~\cite{Julia2017}, if a background gas for which the loss rate coefficients are accurately known and a new gas were flowed through the same orifice flow standard at the same added number density and temperature, the CAVS could determine a ``relative sensitivity coefficient'' between the two gases. 
In such a procedure, the number density divides out and the CAVS retains its primary nature for the new gas.

\subsection{Method IV: Cross species comparisons}
\label{sec:cross-species}

Cross-species comparisons of ratios of loss rates $\Gamma_{\rm loss}$ of co-located lithium and rubidium based CAVSs connected to the same vacuum chamber finds observables~\cite{Shen_2023,frieling2023crosscalibration}
\be
R_{A,B} = \frac{\svtot_{^A{\rm Li}+X}}{\svtot_{^B{\rm Rb}+X}}
\ee
in the limit of ideal CAVSs with trap depth approaching zero. Here, $A$ and $B$ are the atomic numbers of isotopes of Li and Rb, respectively, and $X$ is the background gas species.
This approach does not require knowledge of the background gas number density and is therefore free of many types of systematic errors inherent in efforts to set a known number density. 

\begin{figure}
  \includegraphics[scale=0.28, trim= 0 10 0 0,clip]{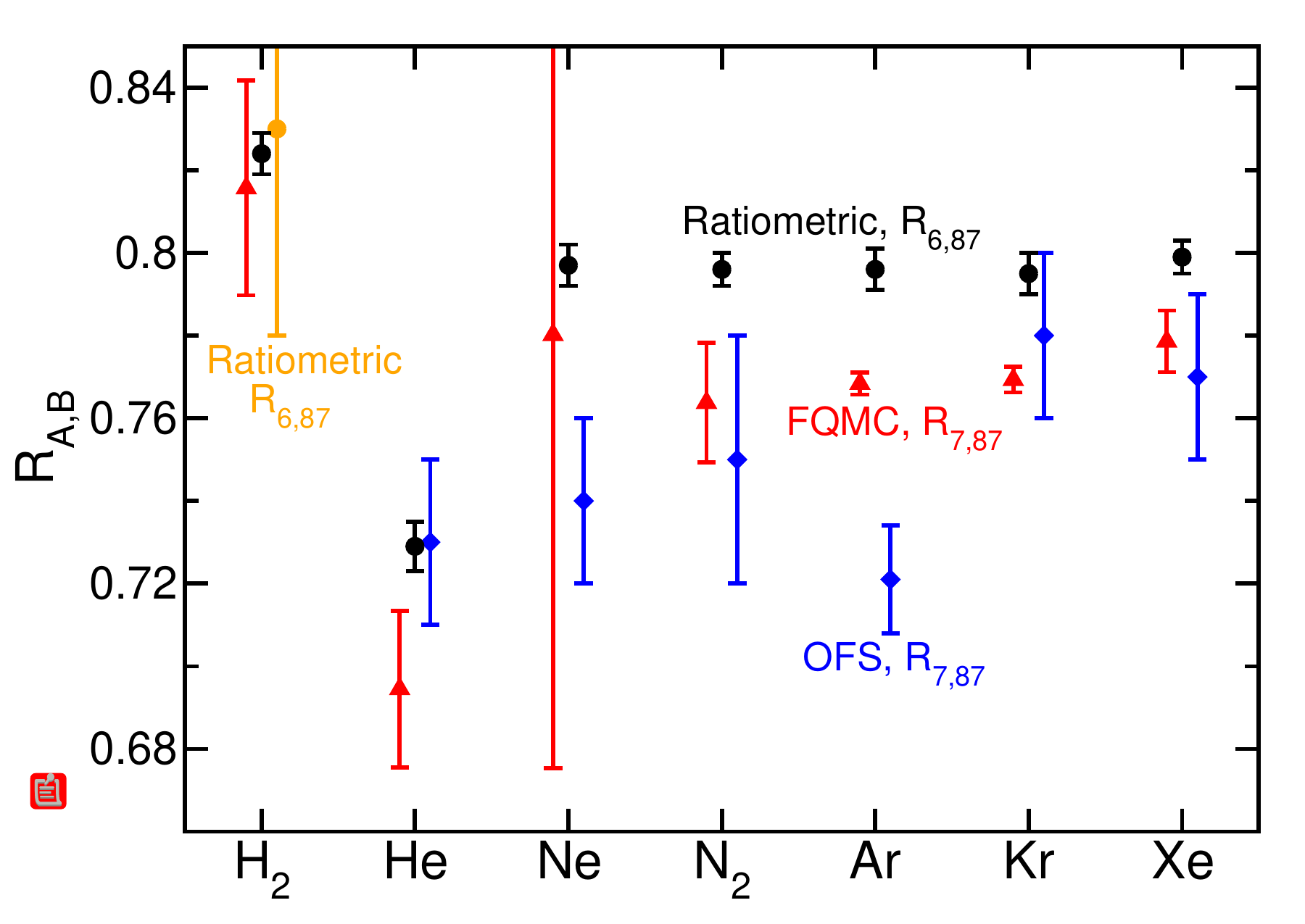} 
\caption{\label{fig:ratio-results} 
Cross-species ratios $R_{A,B}$ of loss rate coefficients from $^A$Li- and $^B$Rb-based cold-atom vacuum sensors for seven background gases at a background gas temperature of $T=298(2)$ K
and in the limit of zero sensor-atom trap depths.
The black filled circles with error bars correspond to $R_{6,87}$
derived from loss rates $\Gamma_{\rm loss}$ for $^6$Li- and $^{87}$Rb-based CAVSs connected to a common vacuum chamber~\cite{frieling2023crosscalibration}.
The filled red triangles and blue diamonds correspond to $R_{7,87}$
from quantum scattering calculations (FQMC) of $\svtot$ for the $^7{\rm Li}+X$ and $^{87}{\rm Rb}+X$ isotopologues \cite{Klos:2023} and  from measurements of $\svtot$ for the same isotopologue pair using an orifice flow standard \cite{Barker2023,eckel_effect_2025}. Finally, the orange filled circle 
shows $R_{6,87}$ for a H$_2$ background gas as measured in Ref.~\cite{Shen_2023}.  
For the black filled circles  standard statistical uncertainties (type A) are specified. For all other data combined systematic and statistical uncertainties are given.
}
\end{figure}

\cite{Shen_2023} performed such ratiometric comparisons with ensembles of co-located $^{6}$Li and $^{87}$Rb sensor atoms exposed to the same H$_2$ gas at $T=300(2)$~K.
In an updated apparatus, \cite{frieling2023crosscalibration} exposed 
co-located $^{6}$Li and $^{87}$Rb sensor atoms to  natural-abundance $X={\rm H}_2$, He, Ne, N$_2$, Ar, Kr, and Xe gases at $T=298(2)$~K.  
Both data sets are shown in Fig.~\ref{fig:ratio-results}.
The figure also shows ratios $R_{7,87}$ {\it inferred} from first-principle quantum calculations with $^7$Li and $^{87}$Rb sensor atoms \cite{Klos:2023} and from experimental work with $^7$Li- and $^{87}$Rb-based CAVSs with known pressures set by an orifice flow standard \cite{Barker2023,eckel_effect_2025}.  

The agreement among the three data sets is satisfactory except for the background gases Ar and Kr,
where either two or all entries differ by more than two standard deviations.
Changes in $R_{A,B}$ due to the use of $^6$Li and $^7$Li isotopes are not the cause of discrepancies. Firstly, \cite{PhysRevA.99.042704,PhysRevA.105.039903} performed calculations of total rate coefficients for the $^6$Li and $^7$Li isotopes colliding
with H$_2$. Their difference between the rate coefficients is 1.5\,\%, slightly less than
their published relative uncertainties. Consequently, the difference between $R_{6,87}$
and $R_{7,87}$ is less than the uncertainty of $R_{7,87}$ from the
first-principle quantum calculations. For the heavier background gases, we can start to
rely on the semi-classical model in Eq.~(\ref{eq:svtot6JB}) for error analysis. This approximation predicts that $R_{A,B}$ is independent of the background gas temperature and of the sensor atom mass, both explicitly and implicitly as to a good approximation the van der Waals $C_6$ coefficient is independent of isotopologues.
{
In any case independent additional measurements and simulations are
required to resolve the discrepancies for Ar and Kr.
}

\section{Applications for a cold-atom vacuum sensor}
\label{sec:application}

There are four practical applications for a CAVS in the UHV domain. 
These are base pressure measurements, measurement of the pressure for some process gas, leak detection, and the calibration of other pressure sensors.
We discuss each of these applications in this section.

\subsection{Base pressure measurement}

The characterization of the lowest pressure that can be obtained in a UHV laboratory is one of the most basic questions to pose and previously not possible to do.
{
See, for example, Fig.~\ref{fig:CAVS_and_other_density-based_methods} showing
the  limited capabilities of ionization gauges in the UHV.
\cite{Ehinger2022}, however,
}
demonstrated just such a measurement of base pressure in their $^7$Li-based CAVS apparatus.
Assuming the background gas was H$_2$, as it is the particle that is hardest to remove or pump from a vacuum chamber, and using the relevant rate coefficient from Table~\ref{tab:Li_summary}, they obtained a relative uncertainty of 2.5~\% on a pressure measurement of $42(1)$~nPa.

\subsection{Process gas pressure measurement}

Using a cold atom pressure sensor to monitor and measure process gases relevant to both industrial and research processes is arguably straightforward, once the relevant $\svloss$ as a function of $U$ and background temperature $T$ is determined via one of the above methods.
That is, measuring the sensor atom number as a function of hold time, thus measuring loss rate $\Gamma_{\rm loss}(U)$ determines $\rn$.
Here, it is worthwhile recalling or repeating some caveats.
Firstly, we cannot accurately calculate or measure the rate coefficients for all process gases.
Still, estimates of rate coefficients with relatively large error bars, say based on Eq.~(\ref{eq:svtot6JB}), can be used.
This limitation also highlights the need for a maintainable database of rate coefficients for all interested parties.
Secondly, industrial processes often operate at medium or high vacuum, where the CAVS cannot operate.
Here, calibration of other pressure-sensing devices with the CAVS will be useful. See also Sec.~\ref{sec:calibration}.

\subsection{Leak detection}

Leaks in large UHV vacuum chambers (such as those employed in gravitational wave detectors and particle accelerators), especially those with irregular shapes and/or multiple flanges and seals, are hard to locate. 
Small leaks allow air, {\it i.e.} N$_2$, O$_2$ and H$_2$O, to enter the vacuum chamber.
The typical procedure to find leaks is to locally spray helium gas on the outside of the vacuum chamber and observe whether and how quickly the pressure inside the chamber increases. 
These observations can then triangulate a leak.
Multiple detectors placed at different locations speed up the process.
{
Currently, the  pressure {\it rise} due to He, but not its value, is detected using species-selective mass-spectrometry systems.
A CAVS device will add  the capability to measure the absolute increase in pressure due to He.
}

\subsection{Calibration of other pressure-sensing devices} \label{sec:calibration}

The use of a CAVS as a pressure standard or reference is of interest for companies that manufacture vacuum sensors or use them to qualify processes/products.
The existing primary ``mechanical'' standards for pressure have to be organized in a chain to connect measurements near atmospheric pressure to those under ultra-high vacuum, as already shown in Fig.~\ref{fig:CAVS_and_other_density-based_methods}.
This means that the measurement standards for pressures near atmospheric pressure, {\it e.g.} mercury manometers, provide traceability to the SI by the quantities of mass, length, and time. 
Measurements sensors for medium and high vacuum are capacitance diaphragm gauges, Fabry-P\'erot refractometry, spinning rotor gauges, and ionization gauges calibrated with static expansion systems.
In these calibrations, known initial pressures are scaled down to much lower pressures by expanding gas from a small volume to a much larger empty volume by knowing the volume ratio.
The initial pressure is measured using mercury manometers.
Ultra-high vacuum standards, {\it i.e.} orifice flow standards, need vacuum gauges calibrated on static expansion systems. 

The measurement chain changes when pressure is realised by cold atom vacuum sensors. The operating principle of these sensors is a direct and independent method and does not depend on standards near atmospheric pressure. As such, they are metrologically attractive. Uncertainties at higher pressures cannot propagate to lower pressures. This means that measuring UHV pressures with cold atoms will verify, improve, and strengthen the  vacuum pressure scale worldwide.

Shen \textit{et al.} \cite{Shen_2020} demonstrated the use of loss of Rb atoms from a magneto-optical trap (MOT) as a transfer standard between a Rb-based CAVS and a spinning rotor gauge using argon as background gas at $T=294$~K.
That is, they determined the loss rate $\Gamma_{\rm{MOT}}$ for Rb atoms in a MOT with a fixed (but unknown) depth for several background pressures $p$ between their base pressure and $10^{-6}$ Pa as measured by their CAVS.  The $10^{-6}$ Pa upper bound corresponded to the largest pressure that their CAVS could accurately detect. 
The slope $S=\Gamma_{\rm{MOT}}/p$ found from a linear fit provides the MOT loss rate coefficient
\be
\svloss_{\rm{MOT}}  =  S\cdot \kb T.
\label{eq:MOTcalibration}
\ee
The MOT was then used as a calibrated pressure gauge capable of measuring argon pressures up to $ 6\times 10^{-5}$~Pa, higher than measurable with their CAVS but reaching the lower range of operation of the spinning rotor gauges.

\section{Other potential applications for a cold-atom vacuum sensor}
\label{sec:other_application}

\subsection{Characterize materials with low outgassing}

The CAVS also holds promise as a tool to better understand the origin of base pressure and what limits the lowest achievable pressure in a vacuum system.
In a leak-free UHV system, the limiting base pressure is determined from the combination of the rate at which gas is removed from the system through pumping and the outgassing rate from the chamber walls.
A recent measurement of outgassing rates from seven geometrically similar but different material chambers showed great variability in both hydrogen and water outgassing~\cite{Fedchak2021}.
Of particular interest were the measurements with titanium.
Titanium outgassing was measured to be one of the lowest and showed no clear Arrhenius-type activation energy when the temperature of the chamber walls was increased.
Reference \cite{Fedchak2021} used a calibrated spinning rotor gauge (SRG) operating between $10^{-4}$~Pa and $1$~Pa.
In principle, an ionization gauge can detect lower pressures than an SRG, but lacks stability and repeatability.
Additionally, ionization gauges produce outgassing and heat that can disturb the measurements.

The CAVS, as a primary standard, is intrinsically stable and be used for pressures at least two orders of magnitude lower than those detectable by an SRG.
As such, a similar experiment with an operable CAVS sensor should have sufficient signal-to-noise to easily measure low outgassing rates and determine effective activation energies.
Thus, for the next generation of vacuum materials, a CAVS can be the ideal choice for measuring outgassing. 

\subsection{Quantum computing/device applications}

A CAVS that operates into the extreme-high vacuum (XHV) regime may prove useful for the development of new quantum devices.
For example, some of the latest experiments with arrays or lattices of Rydberg atoms have realized XHV level vacuums at cryogenic background temperatures with loss rates $\Gamma_{\rm loss}<10^{-3}$~s$^{-1}$~\cite{Schymik2021}.
Such extreme vacuums or low single-body loss rates are currently only obtainable in cryogenic environments, but some experiments are attempting to realize similar loss rates in more-desirable room temperature vacuum chambers.
An operational CAVS  could enable high-throughput testing of vacuum chambers to see if they achieve such low pressures before building a large-scale quantum devices.

\subsection {Benchmarks for molecular scattering theory} 

Measurements of trap loss in the high-vacuum and ultra-high-vacuum regimes provide a means to benchmark quantum chemistry calculations of molecular interaction potentials and molecular scattering theory at few-percent level. In fact, this research field can be viewed as complementary to experiments with molecular beams and experiments on ultracold atom-atom and atom-molecule collisions.  
Molecular beam experiments generally yield energy-resolved differential collision cross sections at large collision angles. Ultracold collision experiments, especially performed as function of applied magnetic field, probe weakly-bound molecular states with binding energies of order $\kb\times1$~mK or less. 

Measurements of sensor-atom trap loss due to background particles at ambient temperatures  probes small-angle scattering processes as a function of trap depth. 
These measurements are then mainly sensitive to long-range van der Waals forces between collision partners, but 
{
also exhibit small, non-trivial sensitivity to the short-range shape of adiabatic Born-Oppenheimer potentials. Nothing is known about the manifestation of the even-smaller non-adiabatic corrections \cite{Agostini2019,Yarkony2023},
such as those due to the  Born-Huang and the recoil corrections. Non-adiabatic corrections are proportional to the electron-to-nuclear mass ratios and expected to be negligible compared to the current uncertainties in the total loss rate coefficients.

Analysis of the anisotropy in the molecular potentials for sensor atom collisions with H$_2$ and N$_2$ has shown that their effect on sensor-atom trap loss is also negligible compared to the uncertainty in the total loss rate coefficients. This might change for more complex molecules like CO$_2$ and H$_2$O.
Recently, \cite{Guo2025}  made progress in answering this question.
}

We believe that new opportunities for sensitive experimental probes into the details of intermolecular interactions lie within these corrections or deviations with promise to reduce the relative uncertainties below one percent. Because the intermolecular interaction potentials for large molecules have complex energy landscapes and because quantum scattering calculations are very time consuming, a meaningful combination of measurements and theoretical calculations may require the development of new theoretical approaches.

\section{Outstanding questions regarding the CAVS as a standard for vacuum metrology}
\label{sec:discrepancies}

Over the next several years, we anticipate research efforts toward resolving some of the outstanding fundamental and technical questions regarding the CAVS.  In this section, we introduce these questions and outline approaches to address them. 

\subsection{What limits do glancing collisions impose?}

The theoretical analysis of the cold-atom vacuum sensor 
{in Sec.~\ref{sec:cavsdetail}}
assumes that the sensor atoms have zero kinetic energy and that glancing-angle collisions with $\theta<\thetamin$ have limited effects on the operation of the sensor.
{
Over time, glancing-angle collisions lead to deviations from the Maxwell-Boltzmann distribution for the sensor atoms and a heated sample, possibly modifying the exponential time evolution of Eq.~(\ref{eq:lossrate}).
}
In order to quantify the effect of the assumptions, analyses have been performed with varying levels of sophistication in Refs.~\cite{Shen_2021,PhysRevA.106.052812,Shen_2023,frieling2023crosscalibration,PhysRevA.109.032818, eckel_effect_2025}.

The authors of Ref.~\cite{PhysRevA.109.032818} developed a  model for the 
time evolution of the kinetic-energy distribution of the sensor-atom ensemble (valid in the limit of no intra-trap collisions) and showed that measurements of the distribution as a function of hold time can be used to determine the total collision rate, $\gtot$, with accuracies below $10^{-3}$, a factor of more than 10 better than previously demonstrated.  
The authors showed that this method would enable a better determination of $\svtot$; however, the analysis requires a calculation of $P(E|E')$, the conditional probability that the sensor atom's kinetic energy post collision is $E$ given that its kinetic energy before the collision is $E'$.  
This conditional probability can be estimated using the universality law. 

Reference~\cite{eckel_effect_2025} applied the model introduced by Ref.~\cite{PhysRevA.109.032818} to correct prior analysis of experimental data to determine $\svtot$ by deriving an analytical approximation for $P(E|E')$ that simplified the analysis of experimental measurements.  
This work confirmed the findings of \cite{PhysRevA.109.032818} that, under certain conditions, the time evolution of the sensor atom number density is no longer exponential.
Neither model accounts for the effects of elastic collisions {\it among} the ultracold sensor atoms. These collisions redistribute kinetic energy among the sensor atoms.

\subsection{What is the best sensor atom?}

The choice of sensor atom for the CAVS affects its operation. If we choose as a reasonable accuracy target for the CAVS of 1\,\%,
the primary difference between $^7$Li and $^{87}$Rb is the difference in the median kinetic energy imparted to the sensor atom in a collision $\ud$ defined by Eq.~(\ref{eq:Udval}). The value for $^{87}$Rb  is about ten times smaller than that for $^7$Li.
Hence, in a typical quadrupole trap with depth $U\sim \kb \times 1$~mK, we anticipate that about 1\,\% of the collisions for $^7$Li impart kinetic energy $E<U$ and about $10$\,\% of the collisions for $^{87}$Rb impart an energy $E<U$. 
Thus, a CAVS using $^7$Li atoms can be modeled as having an idealized trap
with (an almost) vanishingly small depth. A CAVS using $^{87}$Rb atoms, however, will require a measurement of trap depth $U$ to correctly use $\svloss$ and account for glancing collisions.

On the other hand, a Rb-based CAVS has the advantage when knowledge of the loss rate coefficient is required of a gas species that has not been previously measured or calculated.
Here, the larger Rb mass yields a smaller value of $U_{\rm d}$ and this makes it feasible to measure $\svloss$ as a function of $U$ and to use the universality hypothesis \cite{Shen_2020, Shen_2021}.
In this case, operation could proceed in a similar way as above, except that decay curves at different $U$ must be measured to extract not just $\rn$ but also $\svtot$.
References~\cite{Shen_2020,Shen_2021} demonstrated such a measurement and found agreement with a NIST-calibrated ionization gauge, with a total relative uncertainty of better than 10~\% for heavy background species.

Another consideration is the vapor pressure of the sensor atom since these atoms must be introduced into the vacuum.  The sensor atoms thus modify the vacuum under test.  Lithium has the great advantage that its vapor pressure at room temperature is below $10^{-16} \; \mathrm{Pa}$, which is a factor of more than $10^{12}$ times smaller than Rb, so that once Li atoms contact the walls (at ambient temperature), they are effectively removed from the vacuum environment.
Because alkali metal atoms are very reactive, readily forming oxides and nitrides that have much lower partial pressures than these metals, residual gases in the vacuum (e.g.~oxygen or nitrogen) can reduce the vacuum load.  The exact impact of introducing sensor atoms into a vacuum under test remains an open question.

\subsection{Over what density/pressure range can a CAVS be used?} 

To date, the reported pressures measured with CAVSs are in the range of $5\times 10^{-9}$~Pa \cite{Barker2023} to $5\times 10^{-5}$~Pa \cite{Shen_2020,Shen_2021}.
We do not expect these to be fundamental limits to the operating range of a CAVS.

Given the finite time required to prepare the CAVS sensor for a measurement (i.e.~to collect and cool a sufficient number of sensor atoms), the fundamental upper pressure is likely limited to values for which the time interval $1/\gloss$ is longer than this preparation time. The exact pressure where this occurs depends on the background gas species and the sensor atom loading rate, but currently occurs at pressures of the order of $10^{-4}$~Pa. A faster sensor-atom initialization mechanism would allow this limit to be relaxed.

The lower measurable pressure with a CAVS is unknown.
This lower limit will likely be determined by  sensor-atom loss mechanisms that do not dependent on background collisions.  
Evidently, if all sensor atoms are lost before a background collision can occur, then nothing can be said about the background density or pressure.  It is well known that sensor atoms held in a quadrupole magnetic trap can suffer Majorana loss or spin-flip loss as the sensor atom passes near the center of the trap, and this could confound a measurement of the collision-induced loss rate. Estimates reported in Ref.~\cite{Eckel_2018} suggest that Majorana loss would become the dominant loss at or below $10^{-9}$~Pa for Li and $10^{-10}$~Pa for Rb.  The exact nature of Majorana loss is not well understood; however, because it is expected to depend on the sensor atom energy inside the trap, it should result in a non-exponential decay of the sensor atom population.  Understanding the details of this evolution might allow for an analysis that removes this effect, and this would enable accurate extraction of the background-gas collisional loss rate well below the limits stated above. On the other hand, Ioffe-Pritchard (IP) magnetic traps do not suffer from these problems, and IP trap lifetimes of Rb atoms of more than 1000 seconds are common, corresponding to a pressure below 2 nPa.

Another important loss mechanism that must be mitigated are collisions between ultra-cold sensor atoms within the trap.  Such collisions can, for example, lead to two-body spin flips, simultaneously removing two atoms from the trap, as well as to three-body recombination, where weakly-bound di-atomic molecules are formed and three atoms are removed. The processes add the terms $-L_2n^2-L_3 n^3$ to the right hand side of Eq.~(\ref{eq:lossrate})
and thus lead to non-exponential time evolution. Here, $L_2$ and $L_3$ are rates describing the two- and three-body loss processes, respectively. Estimates based on typical sensor-atom number densities below $10^{10}$ cm$^{-3}$ suggest that at least an order of magnitude lower pressures should be detectable~\cite{Barker2022}. Operating at smaller initial sensor-atom number densities while keeping the same initial sensor-atom number will reduce the role of these two effects.

There may exist other considerations that limit the lower range of a CAVS.
In particular, the sources of the sensor atoms are ovens or alkali-metal dispensers.
These devices heat solid alkali metal to sublimate small amounts of atoms into the vacuum chamber for subsequent laser cooling and trapping.
This heating can cause outgassing from the chamber walls and contaminate the vacuum.
Such technical considerations must be better characterized. 

\subsection{What theoretical accuracy for rate coefficients can be achieved?}\label{sec:accuracy} 

The reported measured rate coefficient with a CAVS have an approximate 2~\% relative uncertainty similar to those found from current theoretical simulations with the exception of rate coefficients for neon. 
From the theoretical point of view, if one could solve the relativistic electron and nuclear motion including hyperfine couplings to the spins of the nuclei, the desired elastic and inelastic rate coefficients would have no uncertainty. 

However, due to the multi-electron nature of the colliders, one has to employ approximations, starting with the Born-Oppenheimer approximation that decouples nuclear and electronic motions and leads to potential energy surfaces that nuclei must follow during the scattering events. These surfaces are found using electronic structure programs, such as MOLPRO~\cite{MOLPRO-WIREs}, Gaussian~\cite{Gaussian16}, or Q-Chem~\cite{Q-Chem}, where  many-electron wavefunctions are approximated as expansions in basis functions composed of Gaussian-type orbitals. A truncation in the basis implies a source of uncertainty in the computed potential energy surfaces and dispersion coefficients that translates into an uncertainty in the thermally averaged loss-rate coefficients.

At the moment, the best of the electronic structure methods is the coupled-cluster method including single and double excitations, but accounting for a triple excitations only perturbatively~\cite{KHW93}.
This coupled-cluster method was used in~\cite{Klos:2023}.
Only scalar relativistic corrections were included by~\cite{Klos:2023}.
The accuracy of these surfaces was estimated from calculations by systematically increasing the number of basis functions and from the size of relativistic, spin-orbit corrections.
The combined estimated uncertainties led to the few percent uncertainties in the thermalized rate coefficients. 

To decrease the theoretical uncertainties, first and foremost, the sizes of the electronic basis sets need to be further increased and spin-orbit effects must be included systematically.
The former is currently computationally unfeasible. Once the computational limits have relaxed and the relative uncertainties in the potential energy surfaces approach the electron-to-proton mass ratio,  non-adiabatic corrections must be accounted for. These corrections, among other things, lead to
small, isotope dependent changes in the  potentials and kinetic energy operators for the atoms.
It is currently unknown how to consistently include  non-adiabatic corrections. We infer that relative uncertainties for thermalized rate coefficients of at most of a few times 0.1\,\% might currently be achievable.

\subsection{What gases can we measure with this device? What can be done with complex species like methane or water?}

As of this writing, CAVSs have been tested against all noble gases, N$_2$, CO$_2$, and H$_2$.
These are not the only gases of interest to vacuum metrology.
More reactive gases appear in vacuum systems, like O$_2$, CO, methane, and other simple hydrocarbons.
Perhaps most notably, H$_2$O is a contaminate gas present in almost all vacuum systems.
\emph{A priori} calculations of the molecular potentials and the subsequent collision dynamics for collisions involving water are expected to be infeasible at an accuracy of a few percent.

Experimentally, if the background gas species is known and is sufficiently heavy, then the universality procedure can be applied to estimate the total rate coefficient.  The primary challenge is to ascribe an uncertainty to the $\svloss$ extracted via this technique.  A few species have been studied and the universality procedure was found to systematically underestimate the total collision rate coefficient by up to 7\,\% \cite{universality_revision}. 

For ionization gauges operating in the high-vacuum or milliPascal regime, relative sensitivity factors were obtained in Ref.~\cite{Bartmess1983} by comparing readings of a capacitance diaphragm gauge (CDG), which measures pressure as force per unit area, with those of an ionization gauge, which measures background gas number density.
The former gauge is independent of gas species, while the latter is very much gas species dependent.
In the experiment of Ref.~\cite{Bartmess1983}, the CDG was used to maintain the same pressure as the gas species was switched, and the ratio of the ionization gauge readings with the new gas species and with nitrogen determined relative gas sensitivity factors.
Seventy four relative gas sensitivity factors were measured.

Reference~\cite{Julia2017} proposed a similar technique for the CAVS in the UHV regime,  replacing the CDG with an orifice flow standard to ensure equal gas pressures.
In practice, such a technique might be difficult to implement as
many gases readily stick to common vacuum chamber materials, which can create unwanted long-term dynamics that affect the measurement.
Moreover, some gases may clog the low conductance leaks used within the flow standards.
Further testing of orifice flow standards is required to delineate these limitations.

\section{Technological and commercial challenges for the CAVS}
\label{sec:implementations}

Most applications for cold atom vacuum sensors impose stringent technical requirements.
In the pressure range in which the CAVS operates, it competes with the Bayard-Alpert ionization gauge and its derivatives.
Hence, the requirements on a CAVS are similar to those of this ionization gauge.
Most of these requirements are driven by the industry and market that is addressed.
The relevant technical challenges that go towards a commercial CAVS product are cost,  robustness, ease of operation,  and maintainability. We describe these challenges in more detail below.

\subsection{Cost estimation}

For a successful commercial product, it is essential to be competitive in price and performance with other products available on the market. Currently, the existing cold atom vacuum sensors are individual developments and unique items.
While their performance is outstanding, costs need to be reduced.
Material costs for such devices are approximately 300,000 USD and their setup and optimization requires years of work of trained physicists.
This results in an estimated market price of more than 600,000 USD at the time of this writing.
Compared to ionization sensors for UHV and XHV applications, which cost approximately 10,000 USD, the cost of a CAVS is not yet competitive.

The cost per unit, however, can be significantly reduced when more than one identical unit is produced.
Typically, this reduces the costs for vacuum chambers by a factor of two to six.
Similar scaling might be achievable for optical and electronic components.
A major cost driver  are laser modules, which typically range between 40,000 and 75,000 USD.

Assuming that the quality of the components is well controlled, the time to assemble a device and bring it into operation could be reduced to less than 5 months.
This would allow for prices of 200,000 USD or below.
Further cost reduction will require in-depth analysis and adaption of the design to identify and avoid cost drivers. 
While costs of this order of magnitude will not allow for use of CAVSs as simple process sensors, they might still be attractive for use as an in-house calibration tool for ionization gauges.

\subsection{Robustness}

For industrial applications, even as a device for in-house calibration, the sensor should be small, transportable and robust.
While designs with reduced size using permanent magnets, gratings and small chamber sizes are already operational, see for example the design by NIST shown in Fig.~\ref{fig:pcavs} \cite{Ehinger2022}, the robustness and transportability may be further improved by adopting methods used in aerospace and field projects in the US, France and Germany.
These aerospace applications have driven the development of an ultracold atom apparatus, including all supporting technologies, that can operate on a sounding rocket under strong vibrations with root mean square amplitudes up to $8 g$  and accelerations up to $17 g$ \cite{becker2018space,phdgrosse}, where $g$ is Earth's gravitational acceleration.
Adopting a similarly rugged design will help the device to tolerate vibrations during shipping and during operation in harsh environments.
An effort by the University of Bremen and PTB is underway on the transfer of technology and experience from aerospace and inertial sensor projects to the CAVSs. 

The sensor should also have a single robust package for the lasers, optics, and electronics, such as being mounted in a standardized rack. This has already been realized for the pCAVS at NIST and at the University of British Columbia, although a further decrease in the overall system size, mass, and power budget is desirable.

\subsection{Operation}

The operation of a CAVS should be as simple as possible and have a high degree of autonomous operation. It should provide a stand-alone solution to vacuum measurement with a minimum of training required for its operation.
As such the ``Materiewelleninterferometrie unter Schwerelosigkeit'' (MAIUS) sounding rocket missions use specially developed software \cite{weps2018,elsen2023} that allows easy control of an atom interferometer experiment via  so-called sequences and graphs. Sequences are sets of parameters for single experiments, whereas graphs are decision trees consisting of multiple sequences.
This allows for fast and autonomous execution of experiments also based on the outcome of previous experiments.
In the future, this software shall be applied to the operation of the CAVS, allowing for automated measurements, and evaluation of decay curves to ease the operation of these devices.

\subsection{Maintainability}

Maintenance of the CAVS is a challenging endeavor, especially with regard to parts within the vacuum system that require trained personnel to replace or repair. However, a CAVS can be designed to operate for many years without the need for maintenance.

The only consumable in the CAVS is the source of sensor atoms.  Typically, a vapor of alkali-metal atoms is provided by an atomic oven.  A well-designed oven filled with as little as 1~g of Rubidium can provide the required emission of $10^{8}$~atoms/second for more than 100~years.

An alkali-metal dispenser (AMD) can be used as an alternative sensor-atom source.
These resistively-heated sources evaporate a relatively small amount of alkali-metal, typically on the order of 100~mg, and can produce similar fluxes of atoms as a conventional oven.
They are relatively easy to replace and relatively inexpensive to manufacture.
AMDs made from 3D printed titanium have also shown excellent outgassing properties.
Such printed AMDs have been used in portable versions of the CAVSs~\cite{Ehinger2022}.

The walls of the CAVS are typically made from metal or glass and have metal seals to connect to the chamber for which the pressure needs to be measured.
These materials do not degrade over time and do not need to be replaced.  Ion-getter pumps are required on a CAVS to maintain vacuum and to protect the atomic sources when not connected to a vacuum under test.  The pumping speed of ion-getter pumps decreases over time and saturates to a finite but still non-negligible value.
The saturation time and speed reduction depend on the pressure in the vacuum system.
However, a saturated pump can still be used as long as its speed is sufficient for the CAVS operation. This leaves only the operational lifetime of the pump as a limiting factor.
At a pressure of $10^{-4}$ Pa, the lifetime of  ion-getter pumps is typically specified by the suppliers to be longer than 80\,000 hours (approximately 9 years).
At lower pressures, the lifetime is significantly longer.
A conservative operation time without maintenance of a decade can be expected for a CAVS. 

\section{Goals for the cold-atom vacuum sensor}\label{sec:future}

We have several recommendations for both short- and long-term goals.  These include work to better understand and quantify the operation of a CAVS and work to understand the limits of theoretical models for collision rate coefficients. 
Specifically, there is a need to improve and simplify the control and detection of the sensor atom ensembles.  Motivations on the theory side are the outstanding disagreements between quantum theoretical predictions and experimental measurements, especially for argon,  the surprisingly large uncertainty of rate coefficients for neon from first-principle quantum calculations, 
and our limited knowledge of the uncertainties in the universality hypothesis. 
Some of the goals have been discussed before. Here, we give
different perspectives on these goals as well as describe goals not yet mentioned.

\subsection{Refining measurement precision and accuracy} 

The accuracy and precision of the cold atom vacuum sensor is fundamentally limited by the determination of the loss rate $\Gamma_{\rm loss}$. 
This loss rate is extracted from measurements of the number of sensor atoms remaining in the magnetic trap at several delay times after the preparation of the sensor atom sample. Fluorescence detection is typically achieved by first recapturing the remaining magnetically trapped sensor atoms in the MOT.  The atomic fluorescence induced by the near-resonant laser light creating the MOT is then collected by an imaging lens and detected (See Fig.~\ref{fig:pcavs}).
This process interrupts the number evolution in the magnetic trap, and for each hold time, the experiment must be started over again.

The statistical uncertainty in $\Gamma_{\rm loss}$ is directly related to the statistical uncertainty $u(N)$ in the the number of sensor atoms $n$.
Ref.~\cite{Barker2023} presented an empirical model for $u(n)$ found from $u^2(n)=(\delta_{\rm  n} n)^2 + \delta_0^2$, which combines fluctuations that are proportional to $n$, including fluctuations in the initial number of sensor atoms, with the minimum number of detectable atoms $\delta_0$.
The minimum detectable sensor atom number sets a limit on the longest useful delay time to a few times $1/\Gamma_{\rm loss}$.
In current implementations of the CAVS, $\delta_{\rm n}\approx 0.05$ and $\delta_0\sim 100$~\cite{Barker2023}.

Optimizing the number of delay times required to accurately extract $\Gamma_{\rm loss}$ is  also important and depends on $u(n)$.
Increasing the number of delay times and the number of measurements at the same delay time reduces the statistical uncertainties in the extraction.
Moreover, multiple delay times are needed to distinguish non-exponential from exponential behavior, especially when aiming for accuracies below 1\,\%.
Non-exponential behavior can be due to glancing collisions heating the sensor atoms, Majorana spin flips in the quadrupole magnetic trap, as well as inelastic collisions among the sensor atoms.
Increasing the number of delay times has the downside of increasing the total measurement time for an accurate pressure reading.
An optimal sampling procedure used for multiple sensor-atom measurements is discussed in \cite{frieling2023crosscalibration}.

Non-destructive readout methods, such as phase contrast imaging~\cite{Andrews1996, Higbie2005, Kohnen_2011} or Faraday rotation imaging~\cite{Gajdacz2013}, might replace fluorescence detection.
However, those ``non-destructive'' methods inherently introduce loss of sensor atoms.
The extra, small sensor atom loss will need to be characterized to maintain accuracy and imposes another limit on the uncertainty of CAVS' pressure measurements as well as on the lowest measurable pressure.
The actual additional atom loss depends on both the sensor-atom trap ({\it i.e.} dipole, magnetic, or MOT) and the non-destructive readout method.

Uncertainties in the temperature of the vacuum chamber are also of concern in pressure measurements.
Specifically, temperature uncertainties can arise due to thermometer calibration, temperature gradients across the vacuum chamber walls, and temperature fluctuations during the measurements.
For example, the measurements from NIST are limited by temperature gradients, see Ref.~\cite{Barker2023}.
Formally, these gradients imply  deviations from the Maxwell-Boltzmann distribution.

Still, assuming that corrections to the Boltzmann distribution can be captured in terms of uncertainties,
temperature dependencies in pressure measurements appear due to the ideal gas law and in the loss rate coefficient.
The former has a linear dependence in $T$, while the latter to good approximation is proportional to $T^{3/10}$ from the semi-classical model of Eq.~(\ref{eq:svtot6JB}).
Hence, the approximately 2~K experimental uncertainties near ambient temperature, quoted in Tables \ref{tab:Li_summary} and \ref{tab:Rb_summary}, lead to approximately 1\,\% relative uncertainties in the pressure.
Achieving relative uncertainties on the order of 0.1~\% on the pressure will require temperature uncertainties on the order of 100~mK.

Finally, extracting the total collision rate from a sensor-atom ensemble using a trap of finite depth is complicated. Both the trap-depth-dependent loss rate and the sensor ensemble kinetic energy distribution inside the trap need to be known.  Measurements of the time evolution of the energy distribution together with simulations of a simplified quantum Boltzmann equations for the kinetic energy distribution have been used to determine collision rates with a precision of 0.2\,\% \cite{PhysRevA.109.032818}.  

\subsection{Development of theory}

At present, the most reliably theoretical predictions are carried out by means of first-principle quantum calculations based on the most accurate electronic potential energy surfaces.
Alternative, numerical theoretical approaches might be useful.

For example, ideas rooted in machine learning can be exploited for exploring the sensitivity of predicted measurement outcomes to different parts of multi-dimensional potential energy surfaces. For such applications, the atomic Schr\"{o}dinger equation can be replaced by a machine learning model, trained either by a set of theoretical calculations or by a combination of theoretical results and experimental measurements. Surrogate models can be trained to infer the difference between the theoretical predictions and experimental measurements, while at the same time learning the response of the scattering observables to all Hamiltonian parameters, including various parts of the underlying potential energy surface as well as non-adiabatic couplings. See Ref.~\cite{Guo2025} for a first in-depth analysis using these tools.

Another possibility to improve accuracy is to develop new numerical propagators for close-coupling equations that offer more favorable scaling with the number of parameters that need to be considered for the present applications. 
This can potentially be achieved with techniques based on time-dependent wave-packet propagation. 

\subsection{Quantifying the limits of the universality hypothesis}

The hypothesis that the total collision rate coefficient is universal holds reasonably well.  The most recent evidence comes from comparisons of fully quantum calculations of total rate coefficients with
those from numerical semi-classical models that include $C_6$, $C_8$, and $C_{10}$ dispersion contributions \cite{universality_revision}.  The semi-classical results agree at the 1\,\% level for Rb+$X$ and Li+$X $, when $X$ is one of the heavy collision background species N$_2$, Ar, Kr, and Xe.  
Although the qualitative features of the universality hypothesis are known, it is an approximation and  a quantitative understanding of the sensitivity of $\svtot$ to changes in potential energy surfaces is still lacking.  
In 2024, a preliminary quantitative study examined the response of the thermally averaged rate coefficients for atom-atom collisions to changes in the interaction potential \cite{PhysRevA.110.063317} complementing data in Ref.~\cite{universality_revision}.  Following this, work was done to examine the invariance of the rate coefficients to changes in the anisotropy of atom-molecule interactions \cite{Guo2025}.

\subsection{Connecting to an emerging standard at higher pressures} 

A long-term goal is to develop connections between the CAVS and other emerging standards that measure number densities.
Specifically, we need to extend the operating range of the CAVSs to larger pressures and that of Fabry-P\'erot (FP) refractometers (FPRs) to lower pressures.
Currently, FPRs used for pressure assessments operate at pressures greater than several mPa. Consequently, the two approaches do not have an overlapping pressure range as shown in Fig.~\ref{fig:CAVS_and_other_density-based_methods}.
In refractometry, number density measurements rely on the observation that the resonance frequency of an optical cavity or etalon changes linearly with background gas number density. 
The proportionality factor includes the index of refraction
of the background gas, which can be accurately calculated for
the relevant gasses leading to expanded relative uncertainties below 10$^{-5}$.
For higher number densities, non-linear frequency dependencies, through so-called virial coefficients, need to be considered.

Pressure-induced FP-cavity deformation must be considered and can be assessed with sufficiently low uncertainties using redundant methodologies. Currently, these  involve the alternating use of two gases with  different but well-known refractive indices (preferably helium and argon) or the simultaneous use of two different wavelengths using dichroic resonator mirrors (e.g., at 633 nm and 1550 nm) or, of course, a combination of these two methods.
Moreover, the FP cavity and gas temperature and temperature differentials need to be assessed at the sub-mK level. This can be achieved, provided that a suitable design with appropriately high thermal conductivity and heat capacity is used for the refractometer ~\cite{Rubin2022}.

The remaining challenge in improving FPRs, however, is the length stability of the optical cavities and the corresponding laser lock. Initial measurements with a novel refractometer built at the Physikalisch-Technische Bundesanstalt using argon gas and state-of-the-art optical frequency standards showed that pressures as low as 10$^{-7}$~Pa
can be measured. This paves the way for direct comparisons between CAVSs and FPRs.
It is important to note that, in contrast to a CAVS, FPRs can only measure pressure changes as they compare resonance frequencies when background gasses are present
with that at base pressure.

\section{Alternative pressure sensors in the UHV} \label{sec:alternativeideas}

The main limitation of a cold-atom vacuum sensor is its long measurement time, of order $1/\Gamma_{\rm loss}$, which corresponds to seconds in the UHV domain and greater than 100~s in the XHV domain.
In the following subsections, we explore several measurement platforms that may offer faster measurement time compared to a CAVS.
These measurement platforms fall into two classes.
The first class uses trapped ions to increase the rate coefficient compared to those of a CAVS.
However, the number of trapped ions is small compared to 
{ the number of neutral atoms}
in a CAVS, leading to modest reductions in measurement time.
The second class offers larger potential improvements in the measurement time, but is not amenable to first-principle simulations of rate coefficients.
The second class includes optomechanical sensors and new ionization gauge designs.

\subsection{Using trapped ions}

Arrays of ultracold atomic ions held in Paul or Penning traps can be used as vacuum sensors.
Typical ion traps, however, have orders-of-magnitude higher trap depths than those of magnetic or magneto-optical traps for neutral atoms~\cite{wineland1998}.
Collisions with background gas species are then mostly glancing collisions and inferring the pressure from the loss of ions requires measurement times of hours or days and is not practical~\cite{Pagano2018, Schwindt2016}.
Trapped-ion-based vacuum sensors therefore rely on another process.
That is, glancing collisions cause detectable ion rearrangements within the trap~\cite{Pagano2018,Aikyo2020}.
A less general approach uses the fact that molecular hydrogen H$_2$ reacts readily with optically-excited ions to form hydrides~\cite{Pagano2018,Obil2019}.
This approach might be useful for measuring base pressure as H$_2$ is likely the last remaining species in the XHV and UHV domains.

Ions polarize neutral atoms and molecules through their electric field, leading to an ion-molecule interaction potential proportional to $1/R^4$ for large separations $R$~\cite{wineland1998}.
The longer range and stronger interaction potential yields significantly larger \(\sigma_{\rm tot}\) compared to those derived for CAVSs, so trapped ion vacuum sensors should be able to achieve faster measurement rates than neutral atom sensors.
Thus far, these ion-molecule elastic cross-sections have been computed semi-classically and should be reasonably accurate~\cite{zipkes2011,Hankin2019}.
Fully quantum-mechanical calculations to establish SI traceability have not yet been performed.
The uncertainties in the inelastic rate coefficients for the reactive formation of hydrides using optically-excited ions are significantly larger~\cite{Obil2019}.
Moreover, uncertainties in measuring the population in the excited state of the trapped ion will, because the collision rate constant depends on the internal state, likely limit the accuracy of a pressure measurement at 10~\%~\cite{Eckel_2018,Obil2019}.

The ion rearrangement method uses either careful control of the trapping potential~\cite{Pagano2018, Aikyo2020} or multiple ionic species~\cite{Hankin2019, Aikyo2020} to produce low energy barriers between two or more distinguishable configurations of ions within the trap.
Recently, a fractional pressure measurement uncertainty of  2~\% has been reported using the rearrangement method operating with only two ions~\cite{Hankin2019}.
However, the detectable collision rate in Ref.~\cite{Hankin2019} is on the same order of magnitude as the collision rate in the portable cold-atom sensors of Ref.~\cite{Ehinger2022}.

The main advantage of a trapped-ion pressure sensor is the potential for real-time readout.
The positions of ions can be continuously monitored with the laser-cooling light that maintains their temperature.
Even when the background gas collision rate is low, the absence of a rearrangement informs us about an upper bound on the pressure.
Trapped-ion sensors may therefore be particularly useful in the XHV regime, where the mean time between collisions \(1/\Gamma> 1\)~hour.
In the same pressure regime, readout schemes for cold-atom sensors are likely to be limited by parasitic loss processes driven by collisions among the cold sensor atoms.

\subsection{Using optomechanical systems}

Another way to increase the size of cross sections is to sense collisions with levitated nanospheres.
When the momentum of the nanosphere is measured near the standard quantum limit, the collisions of individual background gas molecules with the nanosphere become detectable~\cite{Ghosh2020, Barker2023b}.
Several experiments have recently demonstrated sensitivity near the standard quantum limit~\cite{delic2020, tebbenjohanns2021} and collisions due to the high momentum tail of the Maxwell-Boltzmann distribution have been measured~\cite{Magrini2021}.
If their results can be extended to smaller, lower mass nanospheres in weaker optical traps, then collisions due to a significant fraction of the Maxwell-Boltzmann distribution will be detectable. 
That is, all changes in momentum $\Delta p>\Delta p_{\rm min}$ with $(\Delta p_{\rm min})^2/(2m_{\rm bg})\ll kT$ are detectable. 

At the same time, we expect $\Delta p_{\rm min} a/\hbar\gg \theta_{\rm qd}$ from practical considerations, where $a$ is the closest approach of a background atom or molecule to a macroscopic, ideal nanosphere and $\theta_{\rm qd} \equiv 2/(k_{\rm rel}a)$, is the quantum diffraction angle for the hard sphere potential defined in Sec.~\ref{sec:history}. Hence, glancing-angle collisions cannot be detected and to good approximation the measured collision rate is given by $\Gamma_{\rm NS}=\rho \langle v_{\rm rel}\rangle \sigma_{\rm cl}$, with collision-energy-independent $\sigma_{\rm cl}$ given by $\sigma_{\rm cl}=\pi a^2$, the classical cross section of a sphere of radius $a$.

For nanospheres suitable for vacuum sensing, the rate coefficient is then roughly 1000 times larger than those for a cold-atom sensor.
Because a nanosphere sensor would combine real-time readout with a large $\langle v_{\rm rel}\rangle\pi a^2$, it could significantly decrease the measurement time.
Realizing a primary pressure measurement with a nanosphere, however, will require control of systematic uncertainties. For example,
the surface area of the nanosphere must be measured \textit{in situ} to better than the desired accuracy of the pressure measurement~\cite{Barker2023b}.
Area uncertainties at the few percent level have been achieved for micron radius nanospheres~\cite{Blakemore2019, Blakemore2020}.
The method of~\cite{Blakemore2019} has yet to be applied to \(100~\si{\nano\meter}\) radius nanospheres that will be necessary for detecting individual background gas collisions.
Efforts using the relevant nanospheres are ongoing~\cite{Carney2020, Afek2022}.
Effects from the nanosphere temperature (often much higher than the background gas temperature), nanosphere surface roughness, and energy transfer to and from the nanosphere  on the collisional cross sections must also be modeled~\cite{Barker2023b}.
It is not yet clear whether a nanosphere based sensor can reach total uncertainties similar to a cold-atom sensor or be operated in industrial environments.

\subsection{Using novel ionization gauges}

The accuracy and repeatability of ionization gauges can be improved if the electron trajectories and kinetic energies are well-defined.
One way to narrow the electron trajectory distribution is to base the ionization gauge design on a focused electron beam~\cite{Jenninger2021}.
Such electron-beam ionization gauges have recently been demonstrated to have linearity of \(0.5\)~\% and reproducibility better than \(1\)~\%~\cite{Jousten2021}.
The sensitivity of electron-beam ionization gauges to nitrogen gas is consistent both from gauge to gauge and with simulations at the \( 2.5\)~\% level~\cite{Jousten2021}.
Relative sensitivities for other gases are also in good agreement with simulations and theoretical calculations~\cite{Jousten2023}.
Electron-beam ionization gauges have measurement uncertainties that are competitive with cold-atom vacuum sensors, but they currently operate from \(10^{-6}~\si{\pascal}\) to \(10^{-2}~\si{\pascal}\).
Whether the design concept and stability of electron-beam ionization gauges can be adapted to the UHV range remains an open question.

\section{Conclusions}\label{sec:conclusion}

We have described the current status of a novel sensor to measure UHV vacuum pressures based on laser-cooled atoms, the cold-atom vacuum sensor. We have shown that existing operational CAVS devices can measure UHV pressures at the few percent level, a significant improvement over state-of-the-art  technologies. In addition, we have shown that CAVSs, unlike all other existing methods, are primary presure sensors as they only require knowledge of calculable collisional rate coefficients between the cold atoms and the atoms and molecules typically present in a UHV vacuum. We have also described the state of the art in calculating these rate coefficients.

To accelerate world-wide dissemination of the cold-atom vacuum sensor, improvements in the design are needed, both to reduce the cost of these devices and to make them easier to use. Refined  calculations of collisional rate coefficients and  improved analysis of the measurement signal would offer relative uncertainties of the pressure below one percent.

{
\section*{Acknowledgment}

Work regarding Fabri-P\'erot refractometry by K.J. and T.R. has been funded by project 22IEM04 MQB-Pascal, which has received funding from the European Partnership on Metrology, co-financed from the European Union’s Horizon Europe Research and Innovation Programme and by the Participating States.

\section*{Data availability statement}

No new data were created or analyzed in this study.

\section*{Ethics statement}

Not applicable.
}

\bibliography{roadmap}
\end{document}